\documentclass[aps,prl,twocolumn,superscriptaddress,showpacs,footnotebib]{revtex4-2}
\usepackage{ulem}
\usepackage{amsmath}
\usepackage{color}
\usepackage{graphicx}
\usepackage{epstopdf}
\usepackage{epsfig}
\usepackage{bm}
\usepackage{amsfonts}
\usepackage[naturalnames]{hyperref}
\usepackage{hypcap}
\usepackage{verbatim}
\usepackage{tabularx}
\usepackage{booktabs}
\usepackage{bbm}
\usepackage{esvect}
\usepackage{tikz}
\usepackage{tikz-feynman}
\usepackage{subfigure}
\usepackage[english]{babel}

\newcommand{\kk}{\mathbf{k}}
\newcommand{\kp}{\mathbf{k}'}
\newcommand{\hc}{\mathrm{H.c.}}
\newcommand{\Do}{\Delta_{\mathrm{orb}}}

\newcommand{\bee}{\begin{eqnarray}}
	\newcommand{\ee}{\end{eqnarray}}
\newcommand{\bma}{\begin{pmatrix}}
	\newcommand{\ema}{\end{pmatrix}}
\newcommand{\balig}{\begin{align}}
	\newcommand{\ealig}{\end{align}}

\newcommand{\ba}{\begin{align}}
	\newcommand{\ea}{\end{align}}

\newcommand{\ignore}[1]{}

\makeatletter

 \graphicspath{{figures/}}
	\makeatother
	
\begin{document}
	\graphicspath{{fig/}}
	\title{Topological superconductivity in superconducting chiral topological semimetals with parallel spin-momentum locking}
	\author{Yingyi Huang}
	\affiliation{School of Physics and Optoelectronic Engineering,
		Guangdong University of Technology, Guangzhou 510006, China}
	\affiliation{Guangdong Provincial Key Laboratory of Sensing Physics and System Integration Applications, Guangdong University of Technology, Guangzhou, 510006, China}
	\date{\today}
	\begin{abstract}
		In contrast to conventional Weyl semimetals in achiral crystals, chiral topological semimetals in chiral crystals exhibit Kramers-Weyl nodes at time-reversal-invariant momenta. A Fermi surface spin texture with parallel spin-momentum locking in these material has been observed by a recent experiment [Nat. Comm. 15,3720(2024)]. We find that the Kramers-Weyl nodes location and the Fermi surface spin texture lead to gapped zero-momenta intranode superconductivity (SC), which is absent in achiral Weyl semimetals. Through self-consistent mean-field calculations, we find that a cubic lattice system in general favors a mixture of spin-singlet $s_\pm$ and $d+id$-wave pairings. In the presence of only the $s_\pm$-wave pairing, we identify a first-order time-reversal invariant topological SC phase. Notably, an SC phase with two Majorana cones for opened Fermi surfaces is energetically favorable. In addition, a second-order topological superconductor with chiral Majorana states can be realized in the presence of a mixture of $s\pm$- and $d+id$-wave pairing. 
		We show that chiral topological semimetals in cubic lattice are fascinating platforms for exploring intrinsic unconventional superconductivity and topological superconductivity.
	\end{abstract}
	\maketitle
	
	\section{\label{sec:intro}I.~Introduction}

	\begin{figure}
		\centering
		\vspace{1mm}
		\hspace{1mm}
		\includegraphics[width=0.95\linewidth]{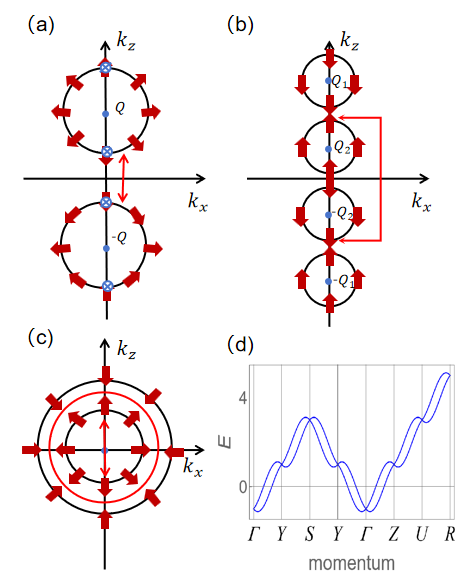}
		\caption{(Color online)(a)-(c) Fermi surface (FS) spin textures around Weyl nodes and Kramers-Weyl nodes in the $k_y=0$ plane and corresponding superconducting pairing states. Arrows indicate the spin components ($\langle\sigma_x\rangle$, $\langle\sigma_z\rangle$) for electron states on FSs. (a) Inversion-symmetric Weyl semimetal in achiral crystal: Cooper pairs form through zero-momentum internode pairing with nodal points (marked by crossed circles) occuring where the spins are maximally antiparallel. (b) Time-reversal symmetric Weyl semimetal in achiral crystal: spin-polarized FSs around Weyl nodes at $Q_1$ ($Q_2$) and $-Q_1$ ($-Q_2$) enable antiparallel spin partners for gapped zero momentum internode pairing. (c) Chiral topological semimetal in chiral crystals: Parallel spin-momentum locking on FSs enable gapped zero-momentum intranode pairing. The red line denotes the $s_\pm$-wave pairing nodal surface separating two Fermi surfaces, realizing first-order topological SC.	
			(d) Band dispersion for a two-band chiral crystal model (Eq.~\ref{Eq:normalH}). 
		}
		\label{fig:fig1}
	\end{figure}		
	Topological superconductors (TSCs) hosting Majorana zero modes have been actively pursued for its important role in fault-tolerant quantum computation~\cite{read2000paired,kitaev2001unpaired,alicea2012new,beenakkerReview,elliott_franz_review,sato2017topological,stanescu2013majorana}. The higher-order counterpart of TSC, which manifests as localized modes at $(d-m)$-dimensional boundaries in a $d$-dimensional material is also an area of intense investigation~\cite{langbehn2017reflection,khalaf2018higher,geier2018second,zhu2018tunable,yan2018majorana,wang2018high,wang2018weak,yan2019higher,liu2018majorana,zhang2019higher,pan2019lattice,zhu2019second,hsu2020inversion,wu2019higher,volpez2019second,wu2020plane,kheirkhah2020first,qin2022topological,wu2020boundary}. However, only a limited number of realistic models have been proposed in intrinsic materials. Very recently, superconductivity have been observed in chiral topological semimetals~\cite{tsvyashchenko2016superconductivity,salamatin2021hyperfine},  a unique class of Weyl semimetals in nonmagnetic chiral crystals. These materials have been considered ideal candidates for conventional TSC~\cite{huang2021three,lee2021topological,mardanya2024unconventional,gao2022topological}. However, the study of higher-order TSC in chiral topological semimetal is still lacking.

	Chiral topological semimetals, realized in crystals lacking chirality-inverting symmetries, have attracted a lot of attention for their host of topological nontrivial fermions~\cite{rao2019observation,takane2019observation,sanchez2019topological,schroter2019chiral,lv2019observation,yao2020observation,schroter2020observation}. These topological fermions, including Kramers-Weyl fermion or multifold fermion, are located at time-reversal invariant momenta (TRIM), i.e., $\textbf{k}_0=-\textbf{k}_0+\textbf{G}$, where $\textbf{G}$ is a reciprocal lattice vector~\cite{bradlyn2016beyond,chang2017unconventional,tang2017multiple}.  This contrasts with Weyl fermions in achiral crystals (which posses mirror or roto-inversion symmetries), where the nodes are located away from TRIM $\textbf{k}_0$. 
	For instance, time-reversal-invariant Weyl semimetals in achiral crystals require at least four Weyl nodes, whereas the minimal number of Kramers-Weyl nodes in a chiral crystal is two.
	In achiral crystals, time-reversal symmetry maps a Weyl node of given chirality at $\textbf{k}\neq\textbf{k}_0$ to another Weyl node of identical chirality at -$\textbf{k}$. This enforces a net zero chiral charge constraint in the Brillouin zone (BZ), necessitating at least four Weyl nodes~\cite{hosur2014time}. In contrast, chiral crystals—lacking such constraints on Kramers-Weyl nodes location—can host Kramers-Weyl nodes exclusively at TRIM, allowing the minimal number of nodes to be two.

	With finite carrier density, Weyl semimetals exhibit disconnected bulk Fermi surfaces (FSs), each surrounds a Weyl node. Figure~\ref{fig:fig1}(a) shows that the inversion-symmetric Weyl semimetal exhibits Weyl node being surrounded by a single FS. Here, the spin states at $\textbf{k}$ and $-\textbf{k}$ can become parallel at the poles of the FS, leading to four nodal nodes in the spin-singlet zero-momentum internode SC state~\cite{cho2012superconductivity,bednik2015superconductivity,PhysRevB.93.094517,PhysRevB.98.024515}. Although a fully gapped SC phase can be realized in the intra-node Fulde-Ferrell-Larkin-Ovchinnikov pairing at finite momentum, the absence of symmetry to relate the paired states precludes a weak-coupling superconducting instability~\cite{bednik2015superconductivity}. In contrast, for time-reversal symmetric Weyl semimetal in achiral crystal (in Fig.~\ref{fig:fig1}(b)), each FS is spin-polarized, potentially facilitating fully gapped internode SC pairing~\cite{hosur2014time,bai2025superconductivity}.

	Compared to Weyl semimetals in achiral crystals with either inversion or time-reversal symmetry, chiral Weyl semimetals exhibit distinct FS configurations and spin textures. In chiral crystals, multiple FSs (two in a two-band model) can surround the same Kramers-Weyl node at TRIM, unlike the single-FS-per-node configuration in achiral systems. {\bf\color{blue}Specifically, in achiral Weyl semimetals, the linear cone dispersion generically yields a single FS per Weyl node, whereas in chiral crystals, the symmetry-enforced non-monotonic band dispersion around Kramers-Weyl nodes at TRIMs robustly produces two concentric FSs~\cite{chang2018topological}.}
	As shown in Fig.~\ref{fig:fig1}(c), these FSs are split by spin-orbit coupling and acquire nontrivial Berry phases from radial spin textures. This radial spin texture, characterized by parallel spin-momentum locking, where the spin is isotropically locked parallel to the electron's linear momentum~\cite{lin2022spin}, is confirmed by spin- and angle-resolved photoemission spectroscopy experiment very recently~\cite{krieger2024weyl}. The separation of FSs creates a large topological non-trivial energy window, leading to helicoid-arc surface states and features intriguing optoelectronic and spintronic properties~\cite{sanchez2019topological,rees2020helicity,dutta2022collective,hsieh2022helicity,ni2021giant}. However, the SC pairing mechanism in chiral Weyl semimetals with this unique FSs configurations and spin-textures features remains an open question.

	The band topology of a gapped odd-parity SC is determined by the relative configuration of FSs and SC pairing surfaces. Chiral topological semimetals feature multiple FSs with parallel spin-momentum locking. It has been proven that an odd number of FSs enclosing TRIM with positive SC pairing leads to a TSC phase~\cite{fu2010odd}. Thus, the even number of FSs in chiral topological semimetals, which carry nontrivial Berry phases, must be separated by SC pairing nodal surfaces to realize a first-order TSC phase. Recently, it has demonstrated that, rather than pairing nodal surfaces, an odd number of removable Dirac pairing nodes enclosed by disconnected FSs are essential for realizing higher-order TSCs~\cite{yan2019higher}. The presence of these removable Dirac nodes guarantees the nontrivial Berry phase on FSs and prevents the continuous contraction of Fermi surfaces to a point or their pairwise annihilation without bulk gap closing.
	
	In this work, we first consider an $s_\pm$-wave SC pairing state with a nodal surface separating the FSs of a chiral Weyl semimetal, yielding a first-order TSC. We next consider the mixed $s_\pm+d_{3z^2-r^2}+i d_{x^2-y^2}$ superconducting state, abbreviated as $s_\pm+d+id$. Hereafter, we denote the real $d_{3z^2-r^2}$ component by $d_q$. This state supports second-order topological superconductivity, with pairing nodal lines located between the normal-state Fermi surfaces without intersecting them. The paper is organized as follows. In Sec.~II, the Weyl semimetals in chiral and achiral crystals are compared. In Sec.~III, a three-dimensional {\bf\color{blue}(3D)} Hubbard model with on-site, nearest neighbor and interband pair-transfer interaction is used to investigate the superconducting pairing instability at the mean-field level. In Sec.~IV, we show the existence of a first-order time-invariant TSC phase with the presence of $s_\pm$-wave SC pairing. In Sec.~V, we present a second-order TSC phases arising from $s_\pm+id$-wave pairing.  
	A further discussion and conclusion are presented in Sec.~VI.

	\section{\label{sec:compare}II. Weyl semimetal in chiral and achiral crystals}
	In this section, we systematically compare the Weyl semimetals in chiral and achiral crystals, focusing on two critical distinctions: A. the locations of Weyl nodes in momentum space which arise from symmetry constraints; B. the spin textures of their Fermi surfaces. Although these two features have been previously proposed for chiral topological semimetals~\cite{chang2018topological}, a comprehensive comparison with Weyl semimetals in achiral crystals has been lacking in the literature. More importantly, these distinctive chracteristic give gapped zero-momentum intranode superconducting pairing in chiral Weyl semimetal.

	\subsubsection{{\bf\color{blue}A. Constraints and Weyl Node Locations}}
	
	Consider a generic two-band Hamiltonian for a spin-$\frac{1}{2}$ system expand around the Kramers-Weyl node:
	\begin{equation}
		\mathcal{H}(\textbf{k})=f_0(\textbf{k})+\bm{f}(\textbf{k})\cdot\bm{\sigma},
		\label{eq:H}
	\end{equation}
	where $\bm{\sigma}=(\sigma_x,\sigma_y,\sigma_z)$ with $\sigma_i$ $(i=x,y,z)$ being Pauli matrices operating on the spin space, $f_0(\textbf{k})$ denotes the kinetic energy term and $\bm{f}(\textbf{k})\cdot\bm{\sigma}$ represents the spin-orbit coupling (SOC) term. The eigenvalues $E_\pm(\textbf{k})=f_0\pm|\bm{f}(\textbf{k})|$ exhibit two-band crossings at Kramers-Weyl nodes when $|\bm{f}(\textbf{k})|=0$.

	In chiral crystals (lacking inversion or roto-inversion symmetry), time-reversal symmetry enforces $\bm{f}(\textbf{k})=-\bm{f}(-\textbf{k})$. Near a TRIM $\textbf{k}_0$, the SOC term must be odd function of $\textbf{k}$. For linear SOC ($\bm{f}(\textbf{k})=\hat{M}\textbf{k}$), the matrix $\hat{M}$ satisfies $det(\hat{M})\neq0$ due to the absence of roto-inversion operations ($R$ with $det(R)=-1$)~\cite{chang2018topological,Xie2021}.
	
	Consequently, {\bf\color{blue}nonmagnetic chiral crystals with SOC can host Kramers--Weyl nodes pinned to TRIM.
		To describe their universal low-energy physics, we consider the minimal two-band lattice Hamiltonian:}
	\begin{equation}
		\label{Eq:normalH}	\mathcal{H}_0(\mathbf{k})=\varepsilon(\mathbf{k})\sigma_0+\sum_{i=x,y,z}\alpha_i\sin k_i \sigma_i,
	\end{equation}
	where  $\varepsilon(\mathbf{k})=-\sum_{i=x,y,z}t_i\cos k_i-\mu$ represents the kinetic energy, and $\alpha_i\sin k_i$ $(i=x,y,z)$ {\bf\color{blue}term represents a Weyl-type SOC. Their momentum dependence is symmetry-enforced by the noncentrosymmetric chiral point group. If the coefficients $t_i$ and $\alpha_i$ are anisotropic, the model belongs to $D_2$ point group. In our discussion, we consider the $O$ point group with the isotropic parameters $t$ and $\alpha$. The interplay between the $\cos k_i$ kinetic terms and the
		$\sin k_i$ SOC terms gives rise to curved, spin-split bands.
		For an appropriate chemical potential, the upper and lower
		helicity branches form the inner and outer Fermi surfaces
		surrounding the same Kramers--Weyl node, respectively, as
		illustrated in Fig.~1(c), consistent with Ref.~\cite{chang2018topological}.}

	In achiral crystals (e.g., with inversion or roto-inversion symmetry), $\det(\hat{M})=0$ due to the existence of at least one mirror or roto-inversion operation $R$ with $det(R)=-1$. This forces at least one SOC component (e.g., $\bm{f}_z$) to  vanish or depend on higher even-order terms on $k$. If $\bm{f}_z=0$, degenerate points form a Kramers nodal line along $k_x=k_y=0$ rather than an isolated point, creating a Kramers nodal line semimetal~\cite{Xie2021}. To obtain a Weyl semimetal, higher-order terms are required. We can consider $\bm{f}_z(\textbf{k})=m-\frac{1}{2}k_x^2-\frac{1}{2}k_y^2$ with $m$ as a mass term, arising from band inversion. The corresponding tight-binding form under the protection of inversion symmetry $\sigma_z\mathcal{H}(\textbf{k})\sigma_z=\mathcal{H}(-\textbf{k})$ ~\cite{cho2012superconductivity} is:
	\begin{equation}
		\begin{aligned}
			\mathcal{H}_I=&t(\sin{k_x}\sigma_x+\sin{k_y}\sigma_y)+t_z(\cos{k_z}-\cos{Q})\sigma_z\\
			&+(2-\cos{k_x}-\cos{k_y})\sigma_z-\mu.	
		\end{aligned}
	\end{equation}
	In the parameter range $0<|\mu/t|\ll Q$, this Hamiltonian describes a time-reversal-broken Weyl semimetal with a single pair of Weyl nodes with different chiral charged located at $k_z=\pm Q$ (Fig.~\ref{fig:fig1}(a)). This differs fundamentally from Weyl semimetals in chiral crystals, where the Kramers-Weyl nodes are constrained to TRIM.
	
	While band inversion mechanism was first proposed for inversion-symmetric topological semimetals, it also applies to time-reversal-symmetric systems~\cite{armitage2018weyl,murakami2008universal}. To achieve nontrivial topology, additional orbital degree of freedom represented by Pauli matrices $\rho_i$ are needed. We can consider a four-band tight-binding model 
	\begin{equation}
		\begin{aligned}
			\mathcal{H}_t&=t(\sin k_y\sigma_z\rho_x-\sin k_x\sigma_0\rho_y)+t_z(\cos k_z-\cos Q)\sigma_0\rho_z\\
			&+(m-\cos k_x-\cos k_y)\sigma_0\rho_z+ t'_z\sin k_z\sigma_z\rho_z-\mu,
		\end{aligned}
	\end{equation} 
	which preserves time-reversal symmetry $\mathcal{T}\mathcal{H}(\textbf{k})\mathcal{T}=\mathcal{H}(-\textbf{k})$ with $\mathcal{T}=-i\rho_y\sigma_0\mathcal{K}$. The term $t'_z\sin k_z\sigma_z\rho_z$ breaks the inversion symmetry and splits the pair of Dirac points at $k_z=\pm Q$ into four Weyl nodes~\cite{hosur2014time}. 
	
	{\bf\color{blue}Crucially, because the Kramers-Weyl nodes in chiral crystals are located at TRIM ($\mathbf{k}_0$), a state at $\mathbf{k}_0+\mathbf{k}$ and its time-reversed partner at $\mathbf{k}_0-\mathbf{k}$ reside within the same node. This unique feature enables the formation of zero-momentum intranode Cooper pairs.}

	\subsubsection{{\bf\color{blue}B. The Fermi surface spin texture}}  
	
	The second distinction manifests in the FS spin texture. While both chiral and achiral Weyl semimetals host Weyl nodes with non-zero chirality, chiral Weyl semimetals exhibit a particularly direct connection between their topological properties and the physical spin texture on the FS. 
	
	In chiral crystals, spin-orbit coupling splits the spin degeneracy everywhere in the Brillouin zone except at the TRIMs [Fig.~\ref{fig:fig1}(d)], creating two time-reversal-symmetric, spin-split FSs at the chosen chemical potential $\mu$ [Fig.~\ref{fig:fig1}(c)]. By definition, the chirality of a Kramers-Weyl node is given by $\mathcal{C}=\text{sign}[v_x\cdot (v_y\times v_z)]$, {\bf\color{blue}where $\mathbf v_i=\nabla_{\mathbf k}f_i(\mathbf k)|_{\mathbf k_0}$
		(with $i=x,y,z$) denote the effective velocities
		~\cite{armitage2018weyl}.
		This topological charge can be equivalently expressed as
		$\mathcal C=\operatorname{sign}[\det M]$, where
		$M_{ij}=\partial f_i/\partial k_j|_{\mathbf k_0}$.
		The same chirality formula applies to achiral Weyl nodes.}
	In chiral topological semimetal, $\boldsymbol\sigma$ represents physical spin,
	and the upper-band spin polarization
	$\mathbf s_{\mathbf k}
	=\langle u_{\mathbf k}|\boldsymbol\sigma|u_{\mathbf k}\rangle$
	is parallel to $\bm f(\mathbf k)$ by the structure of
	$\mathcal H(\mathbf k)$.
	This texture becomes radial in the isotropic linear limit.
	Related spin--momentum locking has been observed by
	spin- and angle-resolved photoemission spectroscopy
	(SARPES)~\cite{krieger2024weyl}.
	Near an isolated linear Kramers--Weyl node, the normalized
	spin polarization $\mathbf s_{\mathbf k}$ sweeps out
	the unit sphere $S^2$ over a closed surface enclosing the node.
	This configuration allows the node chirality to be directly
	extracted from the upper-band spin texture as
	$\mathcal C=\operatorname{sign}(k)\,
	\operatorname{sign}\!\left[
	\mathbf s_{(k,0,0)}\cdot
	\left(\mathbf s_{(0,k,0)}\times\mathbf s_{(0,0,k)}\right)\right]$,
	with momenta measured relative to the node and $k\neq0$
	~\cite{chang2018topological}.
	This expression assumes a locally linear $\bm f(\mathbf k)$,
	while the full band dispersion may remain nonlinear.
	
	In the inversion-symmetric achiral model $\mathcal H_I$,
	the FSs enclosing the Weyl nodes exhibit orbital pseudospin
	textures [Fig.~\ref{fig:fig1}(a)].
	{\bf\color{blue}Here, the Chern number is encoded in the orbital pseudospin
		texture rather than directly in the physical-spin texture.
		Spatial inversion is represented by $P=\sigma_z$, which
		reverses the transverse pseudospin components while preserving
		the Berry curvature at inversion-related momenta,
		$\boldsymbol\Omega(-\mathbf k)=\boldsymbol\Omega(\mathbf k)$.
		Since inversion also reverses the outward normals of the FSs,
		the two inversion-related pockets carry opposite Chern numbers.}
	
	In the time-reversal-symmetric achiral model $\mathcal H_t$,
	the FSs display spin polarizations with antiparallel
	orientations at $\pm k_z$ [Fig.~\ref{fig:fig1}(b)],
	{\bf\color{blue}while the Chern number is encoded in the orbital pseudospin
		texture within each spin block.
		Within each isolated, nondegenerate pocket, all states
		share the same physical-spin eigenstate.
		
		These textures also enter the superconducting pairing projection.
		In chiral crystal, time reversal enforces
		antiparallel physical-spin polarizations at opposite momenta
		relative to the TRIM [Fig.~\ref{fig:fig1}(c)].
		These time-reversal partners lie on the same pocket and allow
		zero-momentum intrapocket spin-singlet pairing, whose band
		projection is analyzed in the following Sec.~IV.
		In $\mathcal H_I$, zero-momentum pairing connects the two
		inversion-related pockets.
		Within the intraband weak-pairing description, their
		Chern-number mismatch requires nodes in the projected gap,
		thereby precluding an everywhere nonvanishing interpocket
		gap~\cite{li2018topological}.
		In $\mathcal H_t$, the fixed physical-spin polarization makes
		the intrapocket spin-singlet projection vanish identically.
		However, pairing between opposite-spin, time-reversal-related
		pockets remains allowed and, for the orbital-independent
		pairing considered here, retains its original magnitude
		under band projection.
		The corresponding pairing projections are derived in the
		Supplemental Material~\cite{suppl1}.
	}

	\section{\label{sec:SC} III.~The relationship between topological superconductivity and \textbf{Fermi surface configuration}}

	As established in the previous section, the Fermi surfaces in chiral topological semimetals exhibit opposite radial spin textures around TRIM. {\bf\color{blue}This feature is compatible with spin-singlet pairing. While our main analysis focuses on spin-singlet pairing, we also assess the effect of the symmetry-allowed $A_1$ spin-triplet admixture through additional self-consistent calculations~\cite{suppl2}. At the representative parameter point examined, the superconducting gaps on the two spin-split FSs remain fully open and retain opposite signs.}
	
	{\bf\color{blue}Under $O$ point group}, the gap functions are reclassified according to its irreducible representations~\cite{smidman2017superconductivity}.
	The $s$-wave pairing transforms under the $A_{1}$ representation. Of particular interest is the anisotropic $s_\pm$-wave pairing, where the superconducting gap changes sign between Fermi surfaces with opposite Chern numbers. This sign reversal is a key ingredient for realizing topological superconductivity, as it can induce nontrivial band topology in the superconducting state. Such sign reversal between the superconducting gaps on different FSs can be achieved through the interplay of Coulomb repulsion and attractive interactions (e.g., phonon-mediated pairing), requiring careful tuning of their relative strengths~\cite{hosur2014time}. Beyond $s$-wave pairing, the even-parity spin-singlet sector also includes the $d_q$ and $d_{x^2-y^2}$ components, which span the two-dimensional $E$ representation.

	To investigate superconductivity, we consider a short-range extended Hubbard
	model on a cubic lattice. In addition to the on-site interaction $U$ and
	the nearest-neighbor density-density interaction $W$, {\bf\color{blue}we include a correlated-hopping interaction $J_{01}$, which describes an
		occupation-dependent modulation of the nearest-neighbor hopping amplitude~ \cite{Hirsch1989,Hirsch2024}.
		Such an effective interaction is allowed in a low-energy description of
		multiple atomic orbitals Kramers--Weyl materials~\cite{suppl3}.
		The interaction Hamiltonian is written as
		\begin{equation}
			\begin{aligned}
				\hat{H}_{\mathrm{int}}
				&=
				U\sum_i n_{i\uparrow}n_{i\downarrow}
				+W\sum_{\langle ij\rangle} n_i n_j
				\\
				&\quad
				+\frac{J_{01}}{4}
				\sum_{\langle ij\rangle,\sigma}
				\left(
				c_{i\sigma}^{\dagger}c_{j\sigma}
				+c_{j\sigma}^{\dagger}c_{i\sigma}
				\right)
				\left(
				n_{i\bar{\sigma}}+n_{j\bar{\sigma}}
				\right),
			\end{aligned}
			\label{eq:Hint}
		\end{equation}
		where $n_i=n_{i\uparrow}+n_{i\downarrow},
		n_{i\sigma}=c_{i\sigma}^{\dagger}c_{i\sigma}$
		and $\bar{\sigma}$ denotes the spin opposite to $\sigma$.
		The notation $\langle ij\rangle$ denotes nearest-neighbor bonds of the lattice.	We adopt the convention that positive $U$ and $W$ denote repulsive onsite and nearest-neighbor density-density interactions, respectively, while negative values denote attraction. The correlated-hopping coupling $J_{01}>0$ mixes the onsite and extended-$s$ pairing components. These couplings are treated as effective low-energy parameters.}

	\begin{figure}
		\centering
		\vspace{1mm}
		\hspace{1mm}
		\includegraphics[width=0.95\linewidth]{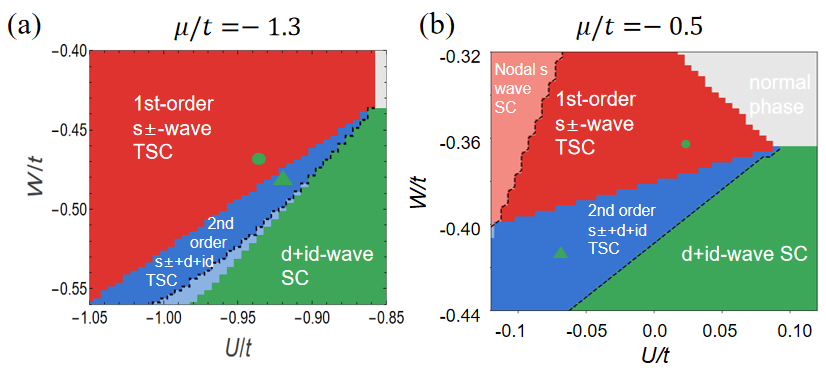}
		\caption{(Color online){\bf\color{blue}Phase diagrams in the $U$–$W$ plane for (a) closed normal-state Fermi surfaces at $\mu/t=-1.3$, $J_{01}/t=0.6$, and $\alpha/t=0.5$, and (b) open normal-state Fermi surfaces at $\mu/t=-0.5$, $J_{01}/t=0.4$, and $\alpha/t=0.8$. Red, blue, and green regions denote the $s$-wave, mixed $s+d+id$, and $d$-wave superconducting sectors, respectively. Here, $s+d+id$ abbreviates $s+d_{3z^2-r^2}+i d_{x^2-y^2}$, and the green region in panel (b) corresponds to a $d+id$ state. The Ginzburg–Landau analysis yields a relative phase of $\pm\pi/2$ between the $s$ and $d_{x^2-y^2}$ components, together with an induced $d_{3z^2-r^2}$ component that is real relative to the $s$ component. The green circle and triangle in each panel mark representative points for first- and second-order topological superconductor, whose spectrum will be shown in figures 3-5. Both panels are calculated at $T/t=0.01$, with $k_B=1$.			
		}}
		\label{fig:fig2}
	\end{figure}

	On the Nambu basis 
	$\Psi^T_\mathbf{k}=(c_{\mathbf{k}\uparrow},c_{\mathbf{k}\downarrow},
	-c_{-\mathbf{k}\downarrow}^\dagger,c_{-\mathbf{k}\uparrow}^\dagger)$, 
	the Bogoliubov-de Gennes (BdG) Hamiltonian is
	$H_\text{BdG}=\frac{1}{2}\sum_\mathbf{k}
	\Psi^\dagger_\mathbf{k}\mathcal{H}_\text{BdG}(\mathbf{k})\Psi_\mathbf{k}$,
	where 
	\begin{equation}
		\begin{aligned}
			\mathcal{H}_\text{BdG}(\mathbf{k})
			&= \left[\varepsilon(\mathbf{k})\sigma_0
			+ \alpha(\sin k_x\,\sigma_x+\sin k_y\,\sigma_y+\sin k_z\,\sigma_z)\right]\tau_z \\
			&- \Delta(\mathbf{k})\tau_x.
		\end{aligned}
		\label{eq:HbdG}
	\end{equation}
	Here $\sigma_i$ and $\tau_i$ are Pauli matrices in spin and Nambu space.
	The superconducting order parameter entering Eq.~(\ref{eq:HbdG}) is
	\begin{equation}
		\Delta(\mathbf{k}) = \Delta_0 f_0(\mathbf{k})
		+ \Delta_s f_s(\mathbf{k})
		+ \Delta_q f_q(\mathbf{k})
		+ i\Delta_d f_d(\mathbf{k}),
		\label{eq:Delta_k}
	\end{equation}
	where $\Delta_0$, $\Delta_s$, $\Delta_q$, and $\Delta_d$ are the amplitudes of the on-site $s$-wave, the extended $A_1$ $s$-wave, the $d_{q}$-wave and the $d_{x^2-y^2}$-wave components, respectively. The $s_\pm$ structure arises from the combined $A_1$ component $\Delta_0f_0+\Delta_sf_s$.
	{\bf\color{blue}The factor $i$ in Eq.~(\ref{eq:Delta_k}) anticipates the relative phase of the $s+d+id$ coexistence state established below.
	}

	{\bf\color{blue}For determining the pairing symmetry, we employ the linearized gap equation
		\begin{equation}
			\Delta_j = -\lambda \sum_{l} \chi_{jl}(T) V_{lm} \Delta_m,
			\label{eq:linearized_gap}
		\end{equation}
		where $\lambda$ is the pairing eigenvalue, $\chi_{jl}$ and $V_{lm}$ are components of the projected pairing interaction and the pairing susceptibility. In the basis $(f_0,f_s,f_q,f_d)$, the projected pairing interaction $V$ is
		\begin{equation}
			V=
			\begin{pmatrix}
				U & J_{01} & 0 & 0\\
				J_{01} & 2W/3 & 0 & 0\\
				0 & 0 & W/3 & 0\\
				0 & 0 & 0 & W
			\end{pmatrix}.
			\label{eq}
		\end{equation}
		Its projections onto the extended-$s$ and two $d$-wave channels are fixed as $2W/3$, $W/3$, and $W$, respectively; they are therefore not independent interaction parameters. The kernel entering the linearized gap equation is $-V$. Whether this gap realizes an $s_\pm$ configuration depends on the FS geometry and is verified explicitly by checking that the gap is nonzero and has opposite signs on the two Fermi surfaces.
		The pairing susceptibility
		\begin{equation}
			\chi_{jl}(T) = \frac{1}{N_k}\sum_{\mathbf{k},n=1,2}
			f_j(\mathbf{k}) f_l(\mathbf{k})
			\frac{\tanh[\xi_n(\mathbf{k})/2T]}{2\xi_n(\mathbf{k})},
			\label{eq:chi}
		\end{equation}
		where
		$\xi_{n}(\mathbf{k})=\varepsilon(\mathbf{k})\pm \alpha(\mathbf{k})$
		with
		$\alpha(\mathbf{k})=\alpha
		\sqrt{\sin^2k_x+\sin^2k_y+\sin^2k_z}$ 
		are the two Rashba-split normal-state bands.The factor 
		$\tanh[\xi_n/2T]/(2\xi_n)$ is the standard BCS thermal weighting. The largest eigenvalue reaches unity at the superconducting transition.

		To determine the relative phase, we perform a Ginzburg-Landau 
		analysis~\cite{suppl4}. Let $\eta_s$, $\eta_1$ and $\eta_2$ denote the complex amplitudes of the leading $s$- and two $d$-wave eigenmodes, respectively. The $s$-wave eigenmode includes both the on-site and extended-$s$ components. Expanding the free energy to quartic order gives
		\begin{equation}
			\begin{aligned}
				F={}&\alpha_s|\eta_s|^2+\alpha_E\rho
				+\beta_s|\eta_s|^4+\beta_1\rho^2+\beta_2|Q|^2\\
				&+\beta_{sE}|\eta_s|^2\rho
				+\gamma\left(\eta_s^{*2}Q+\text{c.c.}\right)\\
				&+\kappa_1\left(\eta_s^* P_1+\text{c.c.}\right)
				+\kappa_2\left(\eta_s^* P_2+\text{c.c.}\right)
				\label{eq}
			\end{aligned}
		\end{equation}
		where all coefficients are real,
		\[
		\rho=|\eta_1|^2+|\eta_2|^2,\qquad Q=\eta_1^2+\eta_2^2,
		\]
		and
		\begin{equation}
			\begin{aligned}
				P_1 &= \eta_1\left(|\eta_1|^2-2|\eta_2|^2\right)
				-\eta_1^*\eta_2^2,\\
				P_2 &= \eta_2\left(2|\eta_1|^2-|\eta_2|^2\right)
				+\eta_2^*\eta_1^2.
			\end{aligned}
		\end{equation}
		where all coefficients are real. The relative phases are determined jointly by the quartic terms. Within the restricted subspace $\eta_1=0$, and for nonzero amplitudes of both $\eta_s$ and $\eta_2$, the $\gamma$ term favors a relative phase of $\pm\pi/2$ when $\gamma>0$, yielding a time-reversal-symmetry-breaking $s_\pm+d_q+id_{x^2-y^2}$ state, and $0$ or $\pi$ when $\gamma<0$, yielding a time-reversal-symmetric $s_\pm+d_q+d_{x^2-y^2}$ state. The existence of a stable coexistence phase therefore requires minimizing the full free energy over all three complex amplitudes and comparing the competing pure and mixed solutions.
		
		We numerically solve the linearized gap equation [Eqs.~(\ref{eq:linearized_gap})-(\ref{eq:chi})] for a range of $U$ and $W$, with $t$ as the energy unit and fixing $J_{01}$ and $\alpha$ separately for each panel. The leading eigenvalues characterize the competing pairing instabilities, and the corresponding eigenvectors determine their form-factor composition. The superconducting phases shown in the phase diagrams are then identified using the Ginzburg--Landau analysis.
		
		As demonstrated in the previous sections, Kramers-Weyl nodes pinned at time-reversal-invariant momenta in chiral crystals can be enclosed by multiple Fermi surfaces. Upon tuning the chemical potential, these Fermi surfaces may undergo a topology change from closed to open as they expand toward and cross the Brillouin-zone boundary. This behavior differs from that in achiral Weyl systems, where Weyl nodes are generally located away from TRIM. In such systems, the Fermi surfaces surrounding distinct Weyl nodes typically merge with one another before reaching the Brillouin-zone boundary, making open Fermi surfaces less common.
		
		To examine how the Fermi-surface geometry influences superconductivity, we consider representative closed- and open-Fermi-surface regimes in Fig.~2. Panel (a), with $\mu/t=-1.3$, corresponds to a closed FS, whereas panel (b), with $\mu/t=-0.5$, represents an open FS. Fig.~2 explores attractive nearest-neighbor interactions ($W<0$) with both attractive and repulsive onsite interactions.

		For closed Fermi surfaces [Fig.~2(a)], the red $s_\pm$-wave phase is favored by more negative $U$, whereas the green $d$-wave sector dominates at weaker on-site attraction. A blue coexistence region lies between them. The coexistence state also contains an induced real $d_q$ component, so its nodal structure is determined from the full complex gap function. We focus below on representative fully gapped states in the dark-red and dark-blue regions, marked by a green circle and triangle, respectively, and examine their first- and second-order topological superconducting properties in the following sections.

		For opened Fermi surfaces [Fig.~2(b)], the superconductor phases persists even for repulsive on-site interactions ($U>0$). The blue $s_\pm+d+id$ coexistence phase extends over a broader range of $W/t$. The red $s_\pm$ phase shinks due to the appearance of nodal $s$-wave sector in light red at more negative $U$ and wider normal phase region.
		
	}

	\begin{figure}
		\centering
		\vspace{1mm}
		\hspace{1mm}
		\includegraphics[width=0.95\linewidth]{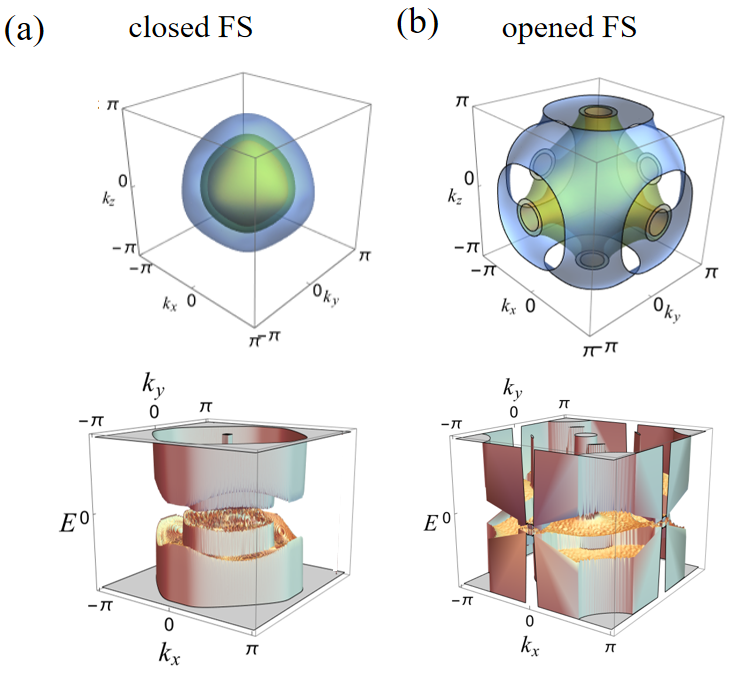}
		\caption{First-order topological superconducting phases measured at the positions labeled by green circle in Fig.~2. Upper panels: Configurations of Fermi surfaces (blue/orange) and pairing nodal surface (green) for (a) closed and (b) open Fermi surfaces. Lower panel: Bogoliubov-de Gennes spectra for slab geometries with open-boundary condition only along $x$. {\bf\color{blue}Parameters: (a) $\Delta_0=-0.009$ and $\Delta_s=0.012$; (b) $\Delta_0=-0.034$, $\Delta_s=0.022$. 
		}}
		\label{fig:fig3}
	\end{figure}

	\section{\label{sec:first}IV.~first-order TSC}
	In this section, we first focus on the regime with $s$-wave SC pairing, in which the $id$-wave SC pairing is small and can be neglect.
	Since the normal part and the $s$-wave pairing part of the Hamiltonian respect time-reversal symmetry (TRS), the BdG Hamiltonian preserves time-reversal symmetry
	\begin{equation}
		\mathcal{T}\mathcal{H}_\text{BdG}(-\mathbf{k})\mathcal{T}=\mathcal{H}_\text{BdG}(\mathbf{k})
	\end{equation}
	associated with the operator $\mathcal{T}=i\tau_0\sigma_y\mathcal{K}$. 
	Also, this model satisfies the particle-hole symmetry  $\Xi\mathcal{H}_\text{BdG}(\mathbf{k})\Xi^{-1}=-\mathcal{H}_\text{BdG}(-\mathbf{k})$ with the operator $\Xi=\tau_y\sigma_y\mathcal{K}$.
	Thus, the system belongs to the symmetry class DIII in 3D, which mathematically classified by an integer invariant $\mathbb{Z}$~\cite{chiu_RMP_16}. They are different from TSC in 2D, which is classified by a $\mathbb{Z}_2$. More importantly, this allows the appearance of a TSC phase with a large topological number.

	It has been demonstrated that an odd-parity pairing gapped superconductor is a first-order TSC if the Fermi surface encloses an odd number of TRIMs in the Brillouin zone~\cite{fu2010odd}. 
	It is easy to see the effective odd-parity pairing from a basis transformation~\cite{suppl5}, after which the BdG Hamiltonian can be decoupled into two independent parts $\mathcal{H}(\bf{k})=\mathcal{H}_+(\textbf{k})\oplus\mathcal{H}_-(\textbf{k})$
	\begin{equation}
		\mathcal{H}_\pm(\bf{k})=\begin{pmatrix}|\varepsilon(\textbf{k})|\pm\alpha(\textbf{k}) & \Delta_\pm(\bf{k})\\
			\Delta^*_\pm(\bf{k}) &-|\varepsilon(\textbf{k})|\mp\alpha(\mathbf{k})
		\end{pmatrix},
	\end{equation}
	where
	$\Delta_\pm(\mathbf{k})=\psi(\mathbf{k})e^{\pm i\varphi_{\mathbf{k}}}$,
	$\psi(\mathbf{k})=\Delta_0+\Delta_s\eta_s(\mathbf{k})+\Delta_q\eta_q(\mathbf{k})$. {\bf\color{blue}Crucially, in the chosen local band gauge, the accompanying phase factor
		\begin{equation}
			e^{\pm i\varphi_{\mathbf{k}}}
			=\frac{\sin k_x\pm i\sin k_y}
			{\sqrt{\sin^2k_x+\sin^2k_y}}
		\end{equation}
		in $\Delta_\pm(\mathbf{k})$ generated by spin-momentum locking gives rise to an effective odd-parity pairing. Its topology is further determined by the Fermi-surface–pairing-node configuration: for two Fermi surfaces, a nontrivial phase occurs when the zeros of \(\psi(\mathbf k)\) separate them, preventing their adiabatic annihilation without gap closing. In contrast, in the achiral models the projected spin-singlet pairing is either nodal on the Fermi surface or vanishes within a fixed-spin pocket, precluding such a fully gapped odd-parity topological phase~\cite{suppl1}.}
	
	In the presence of spin-orbit coupling, the number of normal state FSs is determined by the energy spectra of $\mathcal{H}_-(\mathbf{k})$, since the energy spectra $\mathcal{H}_+(\mathbf{k})$ are always gapped. The FS number must be even due to the time-reversal symmetry. 
	In particular, when $|\varepsilon(\textbf{k})|-\alpha(\mathbf{k})=0$ have solutions, we can see that the normal state of a superconductor has two FSs. Although the $\alpha(\textbf{k})$ has a different form from the general 2D model with Rashba and Dresselhaus spin-orbit couplings~\cite{kheirkhah2020first}, it only changes the shape of Fermi surfaces. 
	
	As demonstrated in Sec.~II, tuning the chemical potential in chiral crystals can drive the Fermi surfaces surrounding Kramers-Weyl nodes at TRIM from closed to open geometries, a situation that is much less common in achiral Weyl systems. To investigate how this distinction affects superconductivity, we consider representative closed- and open-Fermi-surface regimes in Fig.~2.

	The corresponding pairing nodal surface (PNS), defined by $\Delta_s(\mathbf{k})=0$ always separates two regions where $\Delta_s$ has opposite signs. Indeed, according to $\cos k_x+\cos k_y+\cos k_z=-\Delta_0/\Delta_s$, the PNS in three dimensions can form either closed or open surfaces, determined by the sign of $\Delta_0$.

	For comparison, we first look at the case with closed FSs, where the dominant superconducting pairing instability is not favored. A closed PNS must lie between two disconnected FSs to ensure opposite SC gap signs, a key feature of a topological superconductor phase. The surface states of BdG Hamiltonian exhibit a Majorana cone at the $(k_y,k_z)=(0,0)$ point, as can be seen in Fig.~\ref{fig:fig3}(a). According to the bulk-edge correspondence, the number of Majorana cones matches the winding number. In the weak-pairing limit, we can use a formula ~\cite{qi2010topological} 
	\begin{equation}
		\nu=\frac{1}{2}\sum_i sgn(\Delta_i)C_i,
		\label{eq:Qi}
	\end{equation}
	where $sgn(\Delta_i)$ denotes the sign of pairing on the $i$th FS and $C_i$ denotes the first Chern number. The signs of pairing on the two FS are opposite, characterizing a first-order TSC phase with $\nu=1$. 
	
	Let us look at the 3D FSs in the open shape in the first BZ, contrasting with the 2D case where FSs are closed and centered at the corner of FS, which is in a close form. Now we can see that two open FSs are separated by an open PNS without any crossings as shown in Fig.~\ref{fig:fig3}(b). The appearance of open FSs is a natural result of a spin-orbit coupling with three components, which splits FS in three dimensions. 
	
	For the open FS case, we analyze the topology of the gapped superconductor in a space of projectors. The projection operator for occupied state is defined as $P_{\bf k}=\sum\limits_{n\in filled}|u_{n\bf k}\rangle\langle u_{n\bf k}|$ with $|u_{n\bf k}\rangle$ is the $n$-th Bloch wave function at ${\bf k}$. In the presence of chiral symmetry, $Q_{\bf k}=2P_{\bf k}-1$ can be further brought into an off-diagonal form $Q_{\bf k}=\begin{pmatrix} 0 & q_{\bf k}\\q_{\bf k}^\dagger &0\end{pmatrix}$. The winding number can be written as~\cite{schnyder2008classification}
	\begin{eqnarray}
		\nu=\frac1{24\pi^2}\int_{T^3} d^3{\bf k}\epsilon^{ijk}{\rm Tr}\left[{q^\dagger_{\bf k}\partial_iq_{\bf k}q^\dagger_{\bf
				k}\partial_jq_{\bf k}q^\dagger_{\bf k}\partial_kq_{\bf k}}\right],
		\label{eq:Windingnumber}
	\end{eqnarray}
	where the integral is over the first Brillouin zone $T^3$.
	
	In the TSC phase featuring open FSs and a open PNS, we obtain a winding number of $\nu=-2$. Our mean-field calculations confirm the dominant superconducting pairing instability corresponding to this $s_\pm$-wave pairing nodal surface with opened shape. The corresponding surface state of BdG Hamiltonian exhibits two Majorana surface cones at $(k_y,k_z)=(\pi,0$) and $(k_y,k_z)=(0,\pi)$ points. 
	This $\nu=-2$ phase is absent in systems with Rashba and Dresselhaus spin-orbit coupling, which have only two components.

	\begin{figure}
		\centering
		\vspace{1mm}
		\hspace{1mm}
		\includegraphics[width=0.95\linewidth]{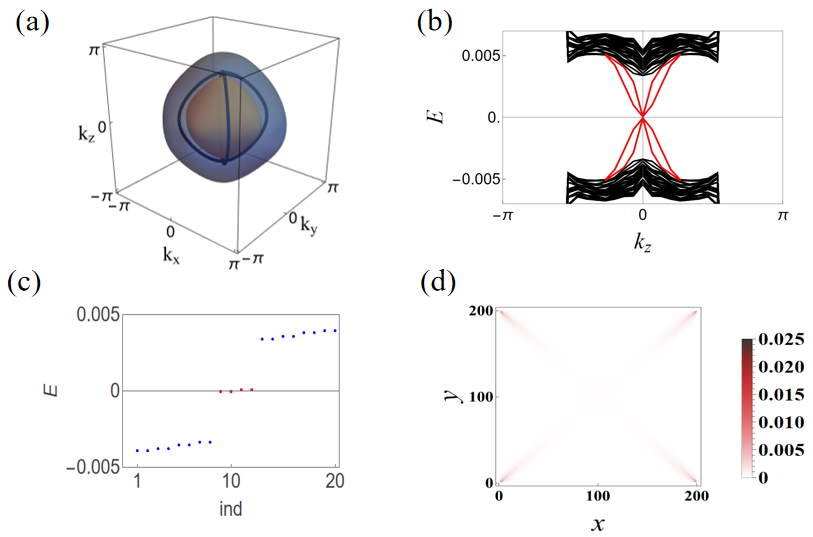}
		\caption{(Color online) Second-order topological superconducting phase for closed FS measured at the positions labeled by green triangle in Fig.~2(a). (a) Configuration of Fermi surface (blue/orange) and pairing nodal lines (black lines satisfying $\Delta_0+\Delta_s\eta_s+\Delta_q\eta_q=0$ and $\Delta_d\eta_d=0$.) (b) Energy spectrum of a cylindrical sample (open boundaries in $x$,$y$) showing chiral Majorana hinge modes crossing at $k_z=0$. Chiral Majorana hinge mode along $z$ direction.  (c) The probability density profiles of four Majorana hinge states and (d) the eigenvalues of the BdG hamiltonian around zero energy (red points). Parameters: $\Delta_0=0.033$, $\Delta_s=0.021$, $\Delta_q=-0.0003$ and $\Delta_d=0.005$.
		}
		\label{fig:fig4}
	\end{figure}
	\begin{figure}
		\centering
		\vspace{1mm}
		\hspace{1mm}
		\includegraphics[width=0.95\linewidth]{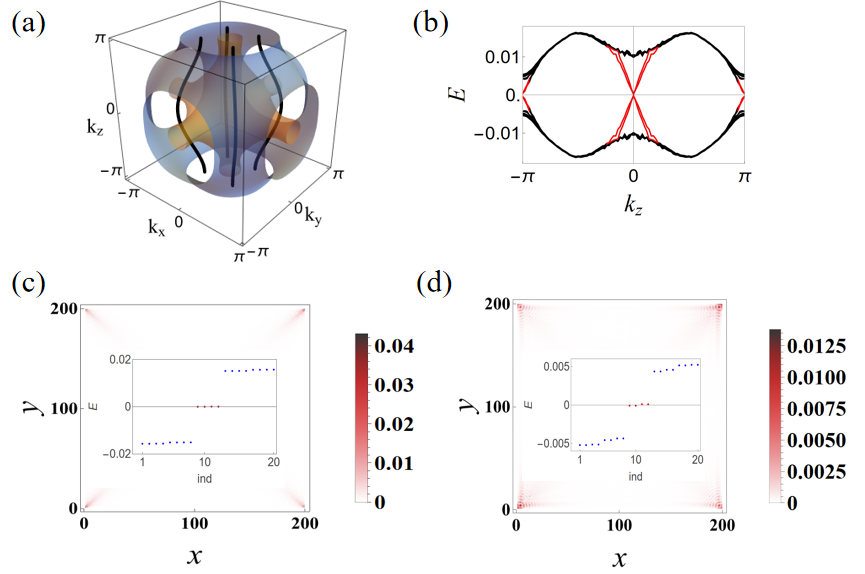}
		\caption{(Color online) Second-order topological superconducting phase for opened FS measured at the positions labeled by green triangle in Fig.~2(b). (a) Configuration of Fermi surface (blue/orange) and pairing nodal lines (black lines satisfying $\Delta_0+\Delta_s\eta_s+\Delta_q\eta_q=0$ and $\Delta_d\eta_d=0$.) (b) Energy spectrum of a cylindrical sample (open boundaries in $x$,$y$) showing chiral Majorana hinge modes crossing at $k_z=0$. Chiral Majorana hinge mode along $z$ direction.  (c) The probability density profiles of four Majorana hinge states and (d) the eigenvalues of the BdG hamiltonian around zero energy (red points). Parameters: $\Delta_0=-0.0014$, $\Delta_s=0.0017$, $\Delta_q=0.0002$ and $\Delta_d=0.0015$.
		}
		\label{fig:fig5}
	\end{figure}
	\section{V.~second-order TSC}
	
	In the section, we investigate the $s_\pm+d+id$ wave SC pairing regime. {\bf\color{blue}We found that the amplitude of the $d_q$-wave SC pairing is very low, at least one orders of magnitude lower than the $s_\pm$-wave component in this mixed pairing state.} Consequently, the $d_q$ has minimal influence on the SC nodal surface. However, the presence of $id$-wave SC breaks the time-reversal symmetry.
	According to the symmetry classification~\cite{chiu_RMP_16}, the 3D superconductor lacking TRS belongs to class D. In this case, the first-order TSC phase becomes trivial, and the helical edge states are gapped out due to TRS breaking.
	
	{\bf\color{blue}The Weyl nodes at the TRIMs endow the normal-state FSs with nontrivial Berry topology, whereas the transition to the second-order TSC is driven by the removable pairing nodal lines rather than by the Weyl nodes themselves.}
	Notably, the $id_{x^2-y^2}$-wave SC order modifies the nodal structure by reducing the pairing nodal surface to four pairing nodal lines along the $k_x=\pm k_y$ directions. These pairing nodal lines occur at generic momenta away from TRIMs, and are therefore removable. When FSs {\bf\color{blue}cross these removable pairing nodal lines} with the bulk gap closing, the system transitions from a first-order to a second-order TSC. The presence of such removable pairing nodal lines separating the two FSs suggests the emergence of a second-order TSC phase, analogous to the removable-pairing-node mechanism discussed in Ref.~\onlinecite{kheirkhah2020first}. While previous studies of pairing nodal points in two-dimensional systems yielded corner states~\cite{kheirkhah2020first}, our three-dimensional system with pairing nodal lines gives rise to hinge states.
	
	The configuration of open FSs and pairing nodal surfaces is shown in Fig.~\ref{fig:fig4}(a). In the first-order TSC phase without $id$-wave SC, we have found two Dirac cones on (100) surface, crossing at $(k_y,k_z)=(0,\pi)$ and $(k_y,k_z)=(\pi,0)$ respectively. When the $id$-wave SC is turned on, we can expect the pairing gaps of the Dirac cones to have sign reversal at $k_z=0$. Figure \ref{fig:fig4}(b) shows that two pairs of chiral Majorana hinge modes cross at $k_z=0$ in the energy spectrum of cylinder sample with open boundary condition in $x$ and $y$ directions. Interestingly, these edge states are isolated from the bulk bands. The spatial profile of the eigenvalues near zero energy confirms the localization of the Majorana hinge modes for $k_z=0$ in Figs.~\ref{fig:fig4}(c-d). The emergence of four non-degenerate Majorana modes - one at each hinge - provides unambiguous evidence of time-reversal symmetry breaking in our system.
	
	{\bf\color{blue}Upon tuning the chemical potential, the FS and pairing nodal surfaces becomes closed, whose configuration is shown in Fig.~\ref{fig:fig5}(a). In the presence of $id$-wave SC, there are chiral Majorana hinge modes exhibit zero-energy crossings at both particle-hole-invariant momenta \(k_z=0\) and \(k_z=\pi\)[Fig.~\ref{fig:fig5}b)]. The spatial profile of the eigenvalues near zero energy confirms the localization of the Majorana hinge modes for $k_z=0$ [Fig.~\ref{fig:fig5}(c)] and for $k_z=\pi$ [Fig.~\ref{fig:fig5}(d)].}

	\section{VI.~discussion and conclusion}
	In this work, we show that both first- and second-order topological superconducting phases can emerge in a chiral topological semimetal. Unlike Weyl semimetals in achiral crystals, chiral topological semimetals are protected by chiral little group symmetry, {\bf\color{blue}leading to Kramers-Weyl nodes located at TRIM}. These Kramers-Weyl nodes are enclosing by Fermi surfaces with radial spin textures, characterized by parallel spin-momentum locking. These unique characteristics particularly favor zero-momentum intranode spin-singlet superconducting pairing. Mean field calculation finds $s_\pm$-wave and $s_\pm+d+id$-wave pairing superconductivity phases. In the presence of $s_\pm$-wave pairing, a first-order TSC phase with $\nu=-2$ emerges naturally from the open Fermi surfaces centered at TRIM. In the presence of $s_\pm+d+id$-wave SC, a second-order TSC phase for open FSs has chiral Majorana hinge modes crossing at two momenta.

	In the above calculations, we use a two-band model in a $O$ point group to show the existence of TSC phases in chiral topological semimetal with Kramers-Weyl nodes. However, our results can be expected to generalize to other chiral crystal materials in cubic lattice with multiple band crossing points, especially, RhGe and AuBe crystallize in the cubic chiral space group \(P2_13\) (No. 198, point group \(T\) which is a subgroup of $O$ point group) and exhibit superconductivity~\cite{salamatin2021hyperfine,matthias1959superconductivity,rebar2019fermi,khasanov2020multiple,khasanov2020single}. Actually, the two features discussed in section II also exist in other chiral topological semimetals. Most importantly, multiple band crossings occur at specific TRIM. Although the increasing number of FS creates a complicated configuration, those FS with distinct Chern numbers near the TRIM at the Brillouin zone center and Brillouin zone corners may still be separated by a pairing nodal surface  if an $s_\pm$-type pairing is realized. For example, in space group 198 chiral topological semimetal, the Fermi surfaces around the $\Gamma$ and $R$ points can be separated by a $s_\pm$-wave pairing nodal surface. Furthermore, the Fermi surface spin textures in other chiral topological semimetals can be more complicated. Although on-site spin-orbital entanglement and the mixing of additional basis states in nonsymmorphic cubic crystals can lead to the vanishing of radial spin-momentum locking when closer to the nodal point~\cite{lin2022spin}, chiral cubic symmetry enforces predominantly radial spin texture with linear momentum dependence on FSs far enough from the nodal point such that bands are only weakly spin-split~\cite{krieger2024weyl,tan2022unified,mera2021different,gosalbez2023diversity}. Furthermore, the normal-state Chern numbers associated with these Fermi surfaces have been confirmed by experiments, indicating that the topological ingredients relevant to our discussion are indeed present~\cite{rao2019observation,takane2019observation,sanchez2019topological,schroter2019chiral,lv2019observation,yao2020observation,schroter2020observation}. Actually, tight-binding model calculation and mean-field analysis have shown the existence of first-order TSC phase with large topological number in chiral topological semimetal with $s_\pm$ SC pairing~\cite{huang2021three}. This suggests that the zero-momentum intranode SC pairing scenario may also apply to chiral topological semimetals in cubic lattices with multiple band crossings, {\bf\color{blue}although a separate symmetry analysis is required for cubic point groups.}
	
	Our analysis of the $s_\pm$-wave and $d+id$-wave pairing coexistence is restricted to cases where the $s_\pm$ component dominates over the $id$ component. However, when the $id$-wave pairing strength becomes comparable to or exceeds that of the $s_\pm$-wave component, the system transitions into a nodal superconducting state. Nevertheless, as demonstrated in the analysis of chiral topological semimetal with multiple band crossing~\cite{mardanya2024unconventional}, such nodal superconductors may also exhibit exhibit topological properties.
	
	Our study extends the topological superconductivity beyond the Rashba spin-orbit coupling physics. In the past decades, the connection between Rashba spin-orbit coupling and Dirac physics in topological materials has been well studied. The spin-momentum locking is admitted as a central ingredient giving rise to remarkable topological properties. However, in recent years, condensed matter physics has found new spin polarization behaviors in Fermi surfaces. For example, in recent studies, the spin texture of the collinear and noncollinear altermagnetism have been linked to topological superconductivity~\cite{lee2024fermi}. We expect the novel spin-momentum locking in Fermi surfaces could give rise to a variety of exotic physics in TSC.

	\section{acknowledgments}
	We would like to thank Zhongbo Yan for helpful discussions. This work is supported by the National Natural Science Foundation of China (Grant No.~12104099 and No.~12274095) and Guangzhou Science and Technology Program (Grant No.~2024A04J0272)
	\bibliography{ref}

\onecolumngrid
\vspace{1cm}
\begin{center}
	{\bf\large Supplemental Material}
\end{center}
\vspace{0.5cm}

\setcounter{section}{0}
\setcounter{secnumdepth}{3}
\setcounter{equation}{0}
\setcounter{figure}{0}
\renewcommand{\theequation}{S-\arabic{equation}}
\renewcommand{\thefigure}{S\arabic{figure}}
\renewcommand\figurename{Supplementary Figure}
\renewcommand\tablename{Supplementary Table}
\newcommand\Scite[1]{[S\citealp{#1}]}
\makeatletter \renewcommand\@biblabel[1]{[S#1]} \makeatother

\section{Superconducting pairing projection}
\subsection{Chiral Weyl semimetal}

In the chosen local band gauge, the off-diagonal pairing
element of each band-resolved BdG block is
\begin{equation}
	\Delta_\pm(\mathbf k)
	=\frac{1}{2}[\Delta_s(\mathbf k)+\Delta_q(\mathbf k)+i\Delta_d(\mathbf k)]
	\frac{\alpha_x(\mathbf k)\pm i\alpha_y(\mathbf k)}
	{\sqrt{\alpha_x^2(\mathbf k)+\alpha_y^2(\mathbf k)}}.
\end{equation}
Here $\Delta_s(\mathbf k)$, $\Delta_q(\mathbf k)$ and $\Delta_d(\mathbf k)$ are even
functions of momentum. Writing
$\hat\Delta(\mathbf k)=\psi(\mathbf k)i\sigma_y$, with
$\psi(\mathbf k)=\Delta_s(\mathbf k)+\Delta_q(\mathbf k)+i\Delta_d(\mathbf k)$,
the intraband projection is
\begin{equation}
	\Delta_\lambda(\mathbf k)
	=\psi(\mathbf k)u_\lambda^\dagger(\mathbf k)
	i\sigma_y u_\lambda^*(-\mathbf k),
	\qquad \lambda=\pm.
\end{equation}
At nondegenerate momenta, its gauge-invariant magnitude satisfies
\begin{equation}
	|\Delta_\lambda(\mathbf k)|^2
	=\frac{|\psi(\mathbf k)|^2}{2}
	\left[1-\hat{\boldsymbol\alpha}(\mathbf k)\cdot
	\hat{\boldsymbol\alpha}(-\mathbf k)\right],
\end{equation}
where $\hat{\boldsymbol\alpha}
=\boldsymbol\alpha/|\boldsymbol\alpha|$.
Time-reversal symmetry of the normal state gives
$\boldsymbol\alpha(-\mathbf k)=-\boldsymbol\alpha(\mathbf k)$,
and hence
\begin{equation}
	|\Delta_\lambda(\mathbf k)|=|\psi(\mathbf k)|.
\end{equation}
Thus, band projection introduces no additional gap nodes.
In particular, the apparent singularities of the phase
representation at $\alpha_x=\alpha_y=0$ with $\alpha_z\neq0$
are gauge artifacts, resolved by changing the local
band-gauge patch.

It is instructive to contrast the above result with two representative achiral systems to demonstrate that the effective odd-parity structure is unique to chiral Weyl semimetals.

\subsection{Inversion-symmetric Weyl semimetal in achiral systems}

For a normal-state basis transformation
$c_{\mathbf k}=U_{\mathbf k}a_{\mathbf k}$, the pairing matrix
in the band basis is
\begin{equation}
	\Delta_{\mathrm{band}}(\mathbf k)
	=U_{\mathbf k}^{\dagger}\hat\Delta(\mathbf k)
	U_{-\mathbf k}^{*}.
	\label{eq:app_pairing_projection}
\end{equation}
We assume an even pairing function,
$\psi(-\mathbf k)=\psi(\mathbf k)$.

Consider inversion-symmetric model
\begin{equation}
	\mathcal H_I(\mathbf k)
	=\xi_0(\mathbf k)\sigma_0
	+\mathbf d(\mathbf k)\cdot\boldsymbol\sigma,
\end{equation}
where $\boldsymbol\sigma$ acts on pseudospin.
Inversion symmetry with $P=\sigma_z$ requires
\begin{equation}
	\mathbf d(-\mathbf k)
	=(-d_x(\mathbf k),-d_y(\mathbf k),d_z(\mathbf k)).
\end{equation}
Let $u_\lambda(\mathbf k)$ denote the normalized eigenstate
of band $\lambda=\pm$, with projector
\begin{equation}
	Q_\lambda(\mathbf k)
	=u_\lambda(\mathbf k)u_\lambda^\dagger(\mathbf k)
	=\frac{\sigma_0+
		\lambda\hat{\mathbf d}(\mathbf k)\cdot\boldsymbol\sigma}{2},
\end{equation}
where $\hat{\mathbf d}=\mathbf d/|\mathbf d|$.
For $\hat\Delta(\mathbf k)=\psi(\mathbf k)i\sigma_y$,
the intraband pairing is
\begin{equation}
	\Delta_{I,\lambda}(\mathbf k)
	=\psi(\mathbf k)u_\lambda^\dagger(\mathbf k)
	i\sigma_y u_\lambda^*(-\mathbf k).
\end{equation}
Using
$\sigma_y\sigma_j^*\sigma_y=-\sigma_j$, its
gauge-invariant magnitude follows as
\begin{align}
	\frac{|\Delta_{I,\lambda}(\mathbf k)|^2}
	{|\psi(\mathbf k)|^2}
	&=
	\operatorname{Tr}\!\left[
	Q_\lambda(\mathbf k)\sigma_y
	Q_\lambda^*(-\mathbf k)\sigma_y\right]
	\nonumber\\
	&=\frac{1-\hat{\mathbf d}(\mathbf k)
		\cdot\hat{\mathbf d}(-\mathbf k)}{2}
	\nonumber\\
	&=\frac{d_x^2+d_y^2}{|\mathbf d|^2}.
	\label{eq:app_inversion_gap}
\end{align}
Thus, even for nonzero $\psi$, the projected pairing vanishes
where $d_x=d_y=0$. At these momenta, the paired states have
parallel pseudospins, so their antisymmetric pairing overlap
is zero.

For $d_x\propto\sin k_x$ and $d_y\propto\sin k_y$, these
conditions must hold simultaneously; neither $k_x=0$ nor
$k_y=0$ alone defines a nodal plane.
A small closed Fermi pocket surrounding a Weyl node on the
$k_z$ axis intersects this axis at two points, producing
two point nodes in the intraband weak-pairing description.

\begin{table*}[t]
	\centering
	\caption{Comparison of zero-momentum pairing in the
		Kramers--Weyl model and the two achiral Weyl models
		$\mathcal H_I$ and $\mathcal H_t$.
		We consider $\hat\Delta=\psi i\sigma_y$ in the two-band
		models and $\hat\Delta=\psi i\sigma_y\otimes\rho_0$ in
		$\mathcal H_t$, with even $\psi(\mathbf k)$.
		Gap statements assume that $\psi$ is nonzero on the
		relevant Fermi surfaces; those for $\mathcal H_I$
		refer to the intraband weak-pairing description.}
	\label{tab:comparison}
	\begin{tabular}{lccc}
		\hline\hline
		& Kramers--Weyl model
		& $\mathcal H_I$
		& $\mathcal H_t$ \\
		\hline
		Node location
		& TRIM & Non-TRIM & Non-TRIM \\
		
		Texture
		& Physical spin $\parallel\pm\boldsymbol\alpha(\mathbf k)$
		& Orbital texture
		& Fixed physical spin $\pm\hat z$ \\
		
		Paired pockets
		& Same TRIM-centered pocket
		& Inversion-related pockets
		& Time-reversal-related pockets \\
		
		Projected pairing magnitude
		& $|\psi|$
		& $|\psi|\sqrt{d_x^2+d_y^2}/|\mathbf d|$
		& 0 in intrapocket \\
		
		Origin of projection zeros
		& None
		& Parallel pseudospins at $d_x=d_y=0$
		& None for opposite-spin partners \\
		
		Gap structure
		& Fully gapped
		& Two point nodes on a small axial pocket
		& Intrapocket singlet pairing forbidden \\
		\hline\hline
	\end{tabular}
\end{table*}
\subsection{Time reversal symmetric Weyl semimetal in achiral systems}
In the basis
$c_{\mathbf k}=(c_{\uparrow1},c_{\uparrow2},
c_{\downarrow1},c_{\downarrow2})_{\mathbf k}^{T}$,
the normal-state Hamiltonian for time-reversal-symmetric model in achiral crystal is
\begin{equation}
	\mathcal H_t(\mathbf k)
	=\begin{pmatrix}
		h_\uparrow(\mathbf k)&0\\
		0&h_\downarrow(\mathbf k)
	\end{pmatrix},
\end{equation}
where $\sigma$ and $\rho$ label physical spin and orbital
degrees of freedom, respectively. Each pocket belongs to
a definite physical-spin sector. The orbital-independent
spin-singlet pairing matrix is
\begin{equation}
	\hat\Delta(\mathbf k)
	=\psi(\mathbf k)i\sigma_y\otimes\rho_0
	=\psi(\mathbf k)
	\begin{pmatrix}
		0&\rho_0\\
		-\rho_0&0
	\end{pmatrix}.
\end{equation}
Let $U_s(\mathbf k)$ diagonalize the orbital Hamiltonian
$h_s(\mathbf k)$. The full normal-state transformation is
\begin{equation}
	\mathcal U_{\mathbf k}
	=\operatorname{diag}
	\left(U_\uparrow(\mathbf k),U_\downarrow(\mathbf k)\right).
\end{equation}
Equation~\eqref{eq:app_pairing_projection} then gives
\begin{equation}
	\Delta_{\mathrm{band}}(\mathbf k)
	=\psi(\mathbf k)
	\begin{pmatrix}
		0&
		U_\uparrow^\dagger(\mathbf k)U_\downarrow^*(-\mathbf k)\\
		-U_\downarrow^\dagger(\mathbf k)U_\uparrow^*(-\mathbf k)&
		0
	\end{pmatrix}.
	\label{eq:app_spin_block_pairing}
\end{equation}
The equal-spin blocks remain identically zero because
$\mathcal U_{\mathbf k}$ does not mix physical spins.
Consequently, spin-singlet pairing within a fixed-spin
pocket vanishes and no pairing phase can be defined in
that channel, thus a helical
factor $e^{\pm i\varphi_{\mathbf k}}$ is absent in the system. This does not preclude azimuthal phase
structure in the normal-state orbital eigenvectors.

Table~\ref{tab:comparison} summarizes how the normal-state
textures affect the specified pairing channels.
In the Kramers--Weyl model, time-reversal partners lie within
the same TRIM-centered pocket, and band projection preserves
the pairing magnitude. In $\mathcal H_I$, the pseudospin
overlap produces zeros in the projected intraband gap.
In $\mathcal H_t$, singlet pairing within a fixed-spin pocket
vanishes, but pairing between opposite-spin,
time-reversal-related pockets preserves $|\psi|$ and can
be fully gapped. 

\section{Effect of spin-triplet pairing}
\label{sec:sm_triplet_pairing}
\begin{figure}[tb]
	\centering
	\includegraphics[width=0.65\linewidth]{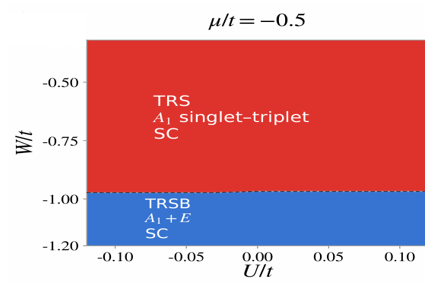}
	\caption{Nonlinear self-consistent $U$--$W$ phase diagram including the
		SOC-aligned $A_1$ triplet component, for $\mu/t=-0.5$, $\alpha/t=0.8$,
		$J_{01}/t=0.4$, and $T/t=0.01$.
		Red and blue denote the time-reversal-invariant $A_1$ mixed phase and the
		time-reversal-breaking $A_1+E$ coexistence phase, respectively.
		The dashed line interpolates the midpoints of the numerical boundary
		brackets; the narrow light band indicates their width.
		The green circle and triangle mark $(U/t,W/t)=(0.025,-0.3725)$ and
		$(0.025,-1.10)$, respectively. The colors classify pairing phases and do
		not assign a higher-order topological invariant.}
	\label{fig:sm_triplet_phase_diagram}
\end{figure}
The absence of inversion symmetry allows spin-singlet and spin-triplet
components to mix when they transform according to the same crystal
representation~\cite{LuYip2008}.
To examine this effect beyond the singlet-only calculation, we include the
SOC-aligned triplet component
$\mathbf{d}(\mathbf{k})=\Delta_t(\sin k_x,\sin k_y,\sin k_z)$.
Although the three orbital basis functions transform as a vector, spin and
momentum must be transformed together in the presence of SOC.
The complete pairing structure is fully symmetric under the cubic $O$ point
group and therefore belongs to $A_1$, allowing it to mix with the $A_1$
singlet component. Thus, spin-triplet character does not by itself imply a
three-dimensional crystal representation.

We retain the four onsite and nearest-neighbor singlet amplitudes and the
additional triplet amplitude, with the projected triplet interaction
$V_{tt}=2W/3$ in our basis convention.
All five amplitudes are allowed to be complex and are determined from the
coupled nonlinear gap equations. Competing solutions are assessed by their
condensation energies and local stability within this pairing space.

Figure~\ref{fig:sm_triplet_phase_diagram} shows the resulting phase diagram
at $\mu/t=-0.5$, $\alpha/t=0.8$, $J_{01}/t=0.4$, and $T/t=0.01$.
The red region is a fully gapped, time-reversal-invariant $A_1$
singlet--triplet mixed state. Stronger nearest-neighbor attraction produces
the blue, time-reversal-breaking $A_1+E$ coexistence region, where the
$E$-channel singlet and the $A_1$ triplet amplitudes are both nonzero.
Full gaps at the sampled points are verified by bounds excluding zeros of
the helicity-resolved gaps on both normal-state Fermi surfaces.
Over $-0.12\leq U/t\leq0.12$, the boundary lies between
$W/t=-0.975$ and $-0.965$.
The sampled solutions in the original parameter window are fully gapped
$A_1$ mixed states, while $A_1+E$ coexistence occurs at stronger attraction.
Triplet pairing therefore substantially modifies the singlet-only phase
diagram, while still permitting time-reversal-breaking coexistence.

At the chosen temperature, the preferred singlet–triplet solutions have lower free energies than the normal state at all sampled parameter points. Therefore, no normal region appears within the scanned window.

The complete helicity-resolved gaps remain nonzero on both Fermi surfaces, with a smallest verified lower bound of approximately \(7.6\times10^{-3}t\). Thus, the sampled states are fully gapped; the absence of nodes is specific to the parameters considered.

\section{ Generation of correlated hopping by orbital downfolding\\in a cubic Kramers--Weyl model}

Occupation-dependent hopping can arise upon low-energy reduction of multiorbital models, as demonstrated for three-band Hubbard models~\cite{Simon1993} and two-orbital dynamic Hubbard models~\cite{Hirsch2003}. Motivated by these constructions, we introduce a symmetry-compatible four-band extension of a Kramers--Weyl model. We first block diagonalize its orbital sector, then project onto the lower doublet and extract the induced correlated hopping. This provides an illustrative microscopic origin of $J_{01}$ rather than a material-specific determination.

\subsection{Parent model and orbital block diagonalization}
Consider two spinful orbitals $a,b$, both transforming as $A_1$ under the chiral cubic point group $O$. In the basis $(a_{\kk\uparrow},a_{\kk\downarrow},b_{\kk\uparrow},b_{\kk\downarrow})^{\mathsf T}$, let
\begin{align}
	\mathcal H_0(\kk)&=\begin{pmatrix}
		h_{\mathrm{KW}}(\kk)&g_{\kk}\sigma_0\\
		g_{\kk}\sigma_0&h_{\mathrm{KW}}(\kk)+\Do\sigma_0
	\end{pmatrix},\\
	h_{\mathrm{KW}}(\kk)&=(-2t\gamma_{\kk}-\mu)\sigma_0
	+\alpha\sum_{\nu=x,y,z}\sin k_\nu\,\sigma_\nu,\\
	\gamma_{\kk}&=\cos k_x+\cos k_y+\cos k_z,\qquad
	g_{\kk}=v_0+2v_1\gamma_{\kk}.
\end{align}
Here $\Do>0$ is the orbital energy offset before hybridization, and $v_0,v_1$ are onsite and nearest-neighbor interorbital hybridizations. All parameters are real. The Hamiltonian is Hermitian and preserves time reversal and $O$ symmetry. The identical dispersive and spin-orbit terms in both orbitals permit an exact orbital rotation:
\begin{equation}
	\begin{pmatrix}a_{\kk\sigma}\\b_{\kk\sigma}\end{pmatrix}
	=R_{\kk}\begin{pmatrix}d_{\kk\sigma}\\e_{\kk\sigma}\end{pmatrix},\qquad
	R_{\kk}=\begin{pmatrix}\cos\theta_{\kk}&\sin\theta_{\kk}\\-\sin\theta_{\kk}&\cos\theta_{\kk}\end{pmatrix},\qquad
	\tan2\theta_{\kk}=\frac{2g_{\kk}}{\Do},
\end{equation}
with $|\theta_{\kk}|<\pi/4$. Setting $\mathcal U_{\kk}=R_{\kk}\otimes\sigma_0$ gives
\begin{align}
	\mathcal U_{\kk}^{\dagger}\mathcal H_0\mathcal U_{\kk}
	&=\operatorname{diag}\!\left(h_{\mathrm{KW}}+\lambda_-\sigma_0,
	h_{\mathrm{KW}}+\lambda_+\sigma_0\right),\\
	\lambda_\pm(\kk)&=\frac{\Do\pm\sqrt{\Do^2+4g_{\kk}^2}}{2}.
\end{align}
This is an exact block diagonalization: all four bands remain, and $d,e$ label the lower and upper orbital sectors. Each block retains its two-component spin structure.

\subsection{Low-energy projection}
Assume that the upper sector is empty and separated from the relevant states, with $\Do$ much larger than the bandwidth, spin-orbit scale, interaction strengths, and $|v_0|+6|v_1|$. Let $P$ project onto the subspace with no $e$ fermions. Writing $u_{\kk}=\cos\theta_{\kk}$, $w_{\kk}=\sin\theta_{\kk}$, and $\eta_{\kk}=g_{\kk}/\Do$, we obtain
\begin{align}
	Pa_{\kk\sigma}P&=u_{\kk}d_{\kk\sigma}
	=\left[1-\eta_{\kk}^2/2+O(\eta_{\kk}^4)\right]d_{\kk\sigma},\\
	Pb_{\kk\sigma}P&=-w_{\kk}d_{\kk\sigma}
	=\left[-\eta_{\kk}+O(\eta_{\kk}^3)\right]d_{\kk\sigma}.
\end{align}
Thus eliminating $e$ does not set $b$ to zero: the retained state contains a small $b$ component. The projected one-particle Hamiltonian is
\begin{equation}
	h_{\mathrm{low}}(\kk)=h_{\mathrm{KW}}(\kk)-\frac{g_{\kk}^2}{\Do}\sigma_0
	+O(g_{\kk}^4/\Do^3).
\end{equation}
The scalar correction renormalizes the dispersion without changing its spin-dependent structure.

Take the microscopic interaction to be
\begin{equation}
	H_{\mathrm{int}}=\sum_i\left(U_a n^a_{i\uparrow}n^a_{i\downarrow}
	+U_b n^b_{i\uparrow}n^b_{i\downarrow}+U_{ab}n_i^a n_i^b\right).
\end{equation}
To first order in these interactions, the effective interaction is $PH_{\mathrm{int}}P$ in the rotated basis; interaction-induced virtual transitions enter at higher orders~\cite{Bravyi2011}. Unlike the strong-coupling reduction in Ref.~\cite{Hirsch2003}, we expand in small orbital mixing and retain terms linear in the interaction strengths.

\subsection{Induced correlated hopping}
Define $r_0=v_0/\Do$, $r_1=v_1/\Do$, and let $\delta$ run over the six nearest neighbors. Since $2\gamma_{\kk}=\sum_\delta e^{i\kk\cdot\delta}$, Fourier transformation yields
\begin{align}
	\widetilde a_{i\sigma}\equiv Pa_{i\sigma}P
	&=(1-r_0^2/2)d_{i\sigma}-r_0r_1\sum_\delta d_{i+\delta,\sigma}
	-\frac{r_1^2}{2}\sum_{\delta,\delta'}d_{i+\delta+\delta',\sigma}+O(r^4),\\
	\widetilde b_{i\sigma}\equiv Pb_{i\sigma}P
	&=-r_0d_{i\sigma}-r_1\sum_\delta d_{i+\delta,\sigma}+O(r^3).
\end{align}
Here $O(r^m)$ denotes total degree at least $m$ in $r_0,r_1$. To identify nearest-neighbor correlated hopping, we isolate the mixed term $r_0r_1$. With
$T_{i\delta,\sigma}=d^\dagger_{i\sigma}d_{i+\delta,\sigma}+\hc$, the density contains
\begin{equation}
	\widetilde a^\dagger_{i\sigma}\widetilde a_{i\sigma}
	=(1-r_0^2)n^d_{i\sigma}-r_0r_1\sum_\delta T_{i\delta,\sigma}
	+\text{terms proportional to }r_1^2+O(r^4).
\end{equation}
The interaction must first be normal ordered, since projection does not generally preserve operator products. In particular,
$n_i^a n_i^b=\sum_{\sigma\sigma'}a^\dagger_{i\sigma}b^\dagger_{i\sigma'}b_{i\sigma'}a_{i\sigma}$.
Substitution of the projected fields gives
\begin{align}
	\left.PH_{U_a}P\right|_{r_0r_1}
	&=-U_a r_0r_1\sum_{i,\delta,\sigma}n^d_{i\bar\sigma}T_{i\delta,\sigma},\\
	\left.PH_{U_{ab}}P\right|_{r_0r_1}
	&=+U_{ab}r_0r_1\sum_{i,\delta,\sigma}n^d_{i\bar\sigma}T_{i\delta,\sigma}.
\end{align}
The interorbital same-spin contributions vanish by fermionic antisymmetry, while $U_b$ starts at fourth order in mixing. Combining the two endpoints of each undirected bond therefore gives
\begin{align}
	H_{\mathrm{ch}}&=X\sum_{\langle ij\rangle,\sigma}
	(n^d_{i\bar\sigma}+n^d_{j\bar\sigma})
	(d^\dagger_{i\sigma}d_{j\sigma}+\hc),\\
	X&=(U_{ab}-U_a)\frac{v_0v_1}{\Do^2}.\label{eq:X}
\end{align}
Each bond is counted once. This term describes hopping modulated by opposite-spin occupation at either endpoint and preserves the parent model's time-reversal and spatial symmetries.

\subsection{Pairing kernel and interpretation of \texorpdfstring{$J_{01}$}{J01}}
Define the singlet Cooper kernel by
\begin{equation}
	H_{\mathrm C}=\frac1N\sum_{\kk,\kp}V_s(\kk,\kp)
	d^\dagger_{\kk\uparrow}d^\dagger_{-\kk\downarrow}
	d_{-\kp\downarrow}d_{\kp\uparrow},
\end{equation}
with both sums over the full Brillouin zone. Exact orbital projection at first order in the interactions gives
\begin{equation}
	V_s(\kk,\kp)=U_a u_{\kk}^2u_{\kp}^2+U_b w_{\kk}^2w_{\kp}^2
	+2U_{ab}u_{\kk}w_{\kk}u_{\kp}w_{\kp}.\label{eq:exactV}
\end{equation}
Expanding to second order in mixing yields
\begin{align}
	V_s(\kk,\kp)={}&U_{\mathrm{eff}}+J_{01}(\gamma_{\kk}+\gamma_{\kp})
	+8U_{ab}r_1^2\gamma_{\kk}\gamma_{\kp}
	-4U_a r_1^2(\gamma_{\kk}^2+\gamma_{\kp}^2)
	+O(U_*\eta_{\max}^4),\\
	U_{\mathrm{eff}}={}&U_a+2(U_{ab}-U_a)r_0^2,\qquad
	J_{01}=4X=4(U_{ab}-U_a)\frac{v_0v_1}{\Do^2}.\label{eq:J}
\end{align}
Here $U_*=\max(|U_a|,|U_b|,|U_{ab}|)$ and $\eta_{\max}=\max_{\kk}|\eta_{\kk}|$.
The coefficient $J_{01}$ multiplies $f_0(\kk)f_s(\kp)+f_s(\kk)f_0(\kp)$ for $f_0=1$, $f_s=\gamma_{\kk}$. Rescaling $f_s$ to $C\gamma_{\kk}$ changes the coefficient to $J_{01}/C$. This is the kernel in the retained spin basis, before any helicity-band projection.

In a phenomenological pairing model, $J_{01}$ mixes onsite and extended-$s$ components; it is not necessary for attraction when $U$ or $W$ is already attractive. For positive diagonal couplings $U,V_{ss}$, a two-channel matrix has an attractive direction if $J_{01}^2>UV_{ss}$. Positive $J_{01}$ favors opposite signs of the components, but an $s_\pm$ state requires checking the gap on both Fermi surfaces.

The microscopic projection also generates the accompanying terms shown above. For $U_a,U_b,U_{ab}\ge0$, its exact first-order kernel, Eq.~\eqref{eq:exactV}, is positive semidefinite, so selective truncation cannot establish a physical attractive channel. Our derivation demonstrates a possible origin of $J_{01}$; attraction from a purely repulsive parent model requires effects beyond this first-order projection.

\section{Ginzburg--Landau Analysis}

We examine the stability of a time-reversal-symmetry-breaking mixed state
within a homogeneous, fourth-order Ginzburg--Landau (GL) description.
For the cubic $O$ point group, the $s$-wave mode belongs to $A_1$, whereas
$d_{3z^2-r^2}$ and $d_{x^2-y^2}$ span the two-dimensional $E$ representation.
We retain the leading $A_1$ eigenmode and both $E$ components in the
spin-singlet sector. The linearized gap equation determines the pairing
modes and their leading instabilities; their equilibrium amplitudes and
relative phases require minimization of the GL free energy.

\subsection{Symmetry-Allowed Free Energy and Microscopic Convention}
Using the model and gap expansion introduced in Sec.~III,
we retain the leading $A_1$ pairing eigenmode and both
components of the $E$ representation. The $A_1$ mode is
described by
\begin{equation}
	\varphi_s(\mathbf{k})
	=u_0f_0(\mathbf{k})+u_sf_s(\mathbf{k}),
	\qquad
	(\Delta_0,\Delta_s)=\eta_s(u_0,u_s),
\end{equation}
where the real coefficients $u_0$ and $u_s$ are obtained
from the linearized gap equation and held fixed during
GL minimization. We use the coefficient convention
$u_0^2+3u_s^2=1$ and omit the second $A_1$ eigenmode.

For the $E$ sector, the normalized basis functions and
amplitudes are related to those in the main text by
\begin{equation}
	\begin{aligned}
		f_1&=-\frac{f_q}{\sqrt{6}},
		&\eta_1&=-\sqrt{6}\,\Delta_q,\\
		f_2&=\frac{f_d}{\sqrt{2}},
		&\eta_2&=i\sqrt{2}\,\Delta_d.
	\end{aligned}
\end{equation}
All three retained amplitudes $(\eta_s,\eta_1,\eta_2)$
are allowed to be complex.

Following the invariant-expansion procedure of Sigrist
and Ueda~\cite{SigristUeda91}, we construct the homogeneous
GL free energy from real scalar combinations invariant
under $O$, global gauge transformations, and time reversal.
Through quartic order, the free-energy density relative
to the normal state is
\begin{equation}
	\begin{aligned}
		\mathcal{F}={}&
		\alpha_s|\eta_s|^2+\alpha_E\rho
		+\beta_s|\eta_s|^4+\beta_1\rho^2+\beta_2|Q|^2\\
		&+\beta_{sE}|\eta_s|^2\rho
		+\gamma\left(\eta_s^{*2}Q+\mathrm{c.c.}\right)\\
		&+\kappa_1\left(\eta_s^*P_1+\mathrm{c.c.}\right),
		\label{eq:GL_free_energy}
	\end{aligned}
\end{equation}
where all coefficients are real and
\begin{equation}
	\begin{aligned}
		\rho&=|\eta_1|^2+|\eta_2|^2,
		\qquad Q=\eta_1^2+\eta_2^2,\\
		P_1&=\eta_1\left(|\eta_1|^2-2|\eta_2|^2\right)
		-\eta_1^*\eta_2^2.
	\end{aligned}
\end{equation}
The decomposition $\operatorname{Sym}^2E=A_1\oplus E$
provides the quadratic combinations used to construct
these quartic invariants. In particular, the $\kappa_1$
term arises from the scalar contraction of
$\eta_s(\eta_1,\eta_2)$ and
$(\eta_1^2-\eta_2^2,-2\eta_1\eta_2)$.
This symmetry-allowed coupling can induce an additional
$d_{3z^2-r^2}$ component, so $\eta_1$ is retained in the
minimization.

The quadratic coefficients are obtained by projecting
the microscopic quadratic pairing kernel onto the retained
$A_1$ mode and the degenerate $E$ doublet. Within the
weak-coupling expansion, the quartic contribution is
\begin{equation}
	\mathcal{F}_4
	=\frac{1}{N_k}\sum_{\mathbf{k},\lambda}
	w_4(\xi_{\lambda\mathbf{k}})
	|\Delta(\mathbf{k})|^4,
	\label{eq:microscopic_quartic}
\end{equation}
where $N_k$ is the total number of momentum points,
$\lambda=\pm$ labels the normal-state bands, and
\begin{equation}
	\begin{aligned}
		w_4(\xi)
		&=\frac{\tanh x-x\operatorname{sech}^2x}{16\xi^3},
		\qquad x=\frac{\xi}{2T},\\
		w_4(0)&=\frac{1}{192T^3}.
	\end{aligned}
	\label{eq:K4}
\end{equation}
Defining
$\langle h\rangle_4=N_k^{-1}\sum_{\mathbf{k},\lambda}
w_4(\xi_{\lambda\mathbf{k}})h(\mathbf{k})$,
we obtain
\begin{equation}
	\begin{aligned}
		\beta_s&=\langle\varphi_s^4\rangle_4,\\
		\beta_1&=\frac{1}{6}\langle f_q^2f_d^2\rangle_4,
		\qquad
		\beta_2=\frac{1}{12}\langle f_q^2f_d^2\rangle_4,\\
		\beta_{sE}&=\frac{2}{3}
		\langle\varphi_s^2f_q^2\rangle_4,
		\qquad
		\gamma=\frac{1}{6}
		\langle\varphi_s^2f_q^2\rangle_4,\\
		\kappa_1&=-\frac{1}{3\sqrt{6}}
		\langle\varphi_sf_q^3\rangle_4.
	\end{aligned}
\end{equation}
These relations follow from the microscopic quartic
functional and the stated basis convention, rather than
from $O$ symmetry alone. The positive weight $w_4$
ensures a positive quartic contribution for a nonzero
gap in the retained pairing space.

\subsection{Amplitude and Phase Stability}

The order parameters are generally complex. For a coexistence state with $\eta_s\neq0$, we fix the overall gauge by choosing $\eta_s=S\geq0$, while allowing both $E$ components to remain complex. We write
\begin{equation}
	\eta_s=S\geq0,\qquad \eta_1=R+ix,\qquad \eta_2=y+iD.
\end{equation}
We minimize Eq.~\eqref{eq:GL_free_energy} over all five real variables
$\mathbf{v}=(S,R,x,y,D)$ using multiple initial conditions, without
imposing a relative phase or setting either $E$ component to zero.
A stationary solution is a strict local minimum when its gauge-fixed
Hessian
\begin{equation}
	H_{ij}=\left.\frac{\partial^2\mathcal{F}}
	{\partial v_i\partial v_j}\right|_{\mathbf{v}_*}
\end{equation}
is positive definite. Its energy is then compared with those of the
normal state, the optimized pure-$s$ and pure-$E$ states, and other mixed
solutions found in the search.

In the restricted subspace $\eta_1=0$, the $\gamma$ term favors
$\eta_2/\eta_s$ purely imaginary for $\gamma>0$. In the full space,
however, the $\kappa_1$ term can induce a real $d_{3z^2-r^2}$ component.
Indeed, for $(\eta_s,\eta_1,\eta_2)=(S,R,iD)$,
\begin{equation}
	\left.\frac{\partial\mathcal{F}}{\partial R}\right|_{R=0}
	=-2\kappa_1SD^2.
\end{equation}
Consequently, a nonzero $\kappa_1$ generally induces a finite $d_{3z^2-r^2}$ component in the coexistence state. When this additional component is small, the resulting state can be approximately described as $s+id$, while its stability must be assessed within the full pairing space.

\subsection{Representative-Point Stability and Phase Scan}

We examine the representative state marked in Fig.~2(b), using
$U/t=-0.065$, $W/t=-0.435$, $J_{01}/t=0.4$, $\mu/t=-0.5$,
$\alpha/t=\alpha_z/t=0.8$, and $T/t=0.01$, with $t_z=t$ and $t'=0$.
The interaction and band-sum conventions are the same as those used
for the representative-state calculation. We retain both the on-site and extended-$s$ components in the
$A_1$ sector, together with the two components of the $E$
representation, allowing all four complex amplitudes to vary independently.

The interaction matrix contains a repulsive $A_1$ channel, so local
stability is assessed using a variational construction rather than
unrestricted minimization of the gap-equation
functional~\cite{Fanfarillo2025}.
Let $\mathbf z$ collect the four gap coefficients in the same real
basis as $V$, and let $\Omega_4(\mathbf z)$ denote the fermionic
free energy relative to the normal state, expanded through fourth
order. We use
\begin{equation}
	\begin{aligned}
		\mathbf p(\mathbf z)
		&=-\frac{\partial\Omega_4}{\partial\mathbf z^*},\\
		\mathcal F_{\mathrm{var}}(\mathbf z)
		&=\Omega_4(\mathbf z)
		+2\operatorname{Re}[\mathbf z^\dagger\mathbf p(\mathbf z)]
		+\mathbf p^\dagger(\mathbf z)V\mathbf p(\mathbf z).
	\end{aligned}
	\label{eq:GL_variational_functional}
\end{equation}
At a self-consistent solution, $\mathbf z+V\mathbf p=0$, its value
agrees with that of the gap-equation functional. This construction
generates induced higher-order terms from $\Omega_4$ and is used
here as a GL-derived approximation for the local stability test.

On a $256^3$ momentum grid, the mixed solution has the form
$s+d_{3z^2-r^2}+id_{x^2-y^2}$, together with its time-reversed
partner. In the convention of Eq.~\eqref{eq:Delta_k}, all four
amplitudes can be chosen real, with nonzero $\Delta_q$ and
$\Delta_d$; the imaginary coefficient of $f_d$ is supplied by
the explicit factor $i$.
The Hessian of $\mathcal F_{\mathrm{var}}$ has seven positive
eigenvalues and one zero eigenvalue associated with the global phase.
The mixed-state, pure-$E$ chiral, and pure-$A_1$ energies are
$-1.1607460\times10^{-4}$, $-1.1227764\times10^{-4}$, and
$-9.3532605\times10^{-5}$, respectively, in the stated normalization.
Thus,
\begin{equation}
	\mathcal F_{\mathrm{var,mix}}
	<\mathcal F_{\mathrm{var},E_{\mathrm{ch}}}
	<\mathcal F_{\mathrm{var},A_1}<0.
	\label{eq:thermodynamic_comparison}
\end{equation}
The mixed-state energy changes by approximately $0.038\%$ between
the $192^3$ and $256^3$ grids. These results support local stability
and an energetic preference over the compared states within the
stated approximation.

\begin{figure}[t]
	\centering
	\includegraphics[width=0.65\linewidth]{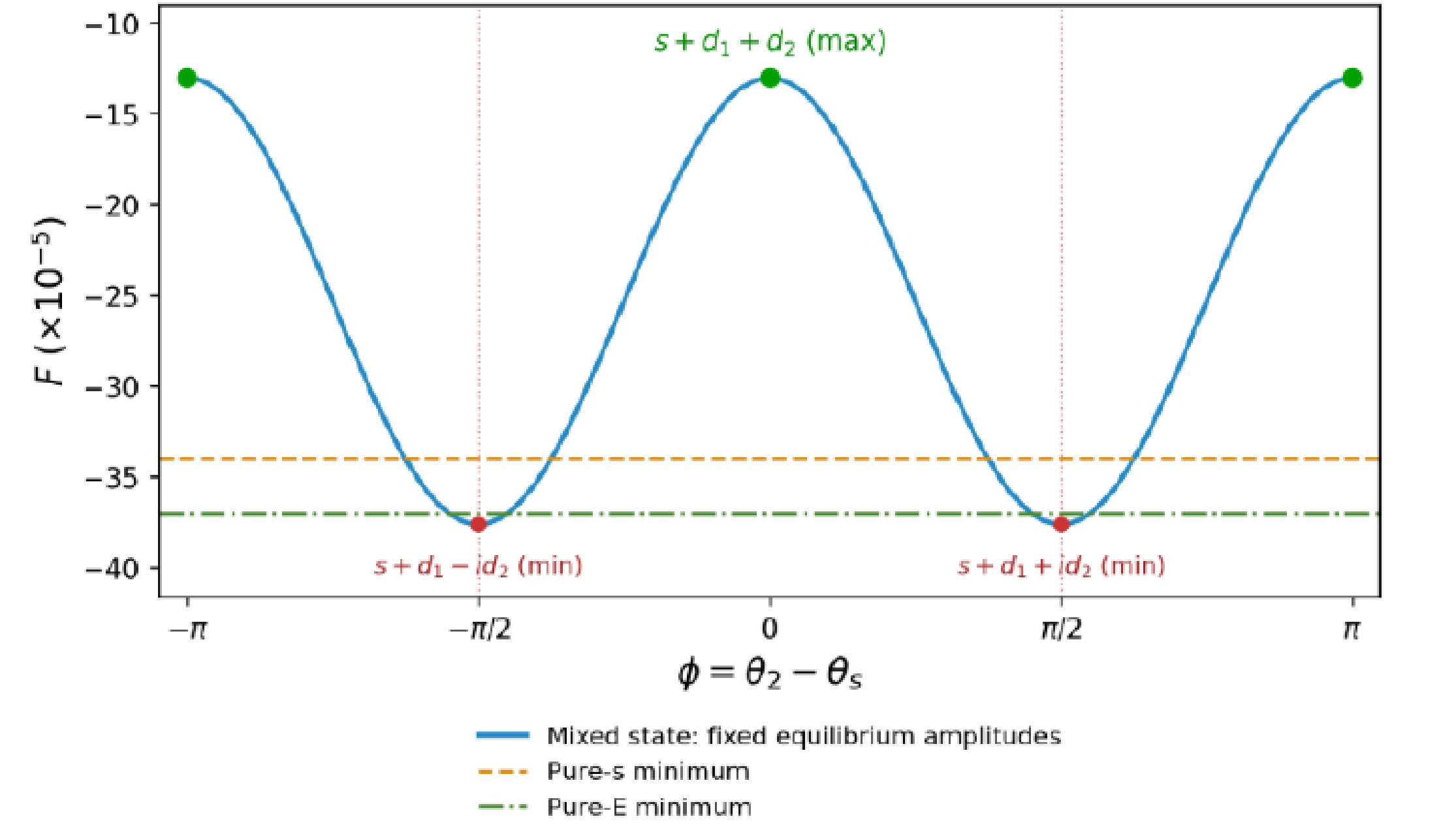}
	\caption{(Color online) Fixed-amplitude phase dependence of the $O$-symmetric $A_1\oplus E$ GL free energy at
		$U/t=-0.065$, $W/t=-0.435$, $J_{01}/t=0.4$, $\mu/t=-0.5$,
		$\alpha/t=\alpha_z/t=0.8$, and $T/t=0.01$,
		evaluated on a $256^3$ momentum grid.}
	\label{fig:feynman_GL}
\end{figure}

To visualize the phase dependence, we fix the amplitudes at their
mixed-state equilibrium values and vary only the phase of the
$d_{x^2-y^2}$ coefficient. With the first three coefficients real,
we write
$\mathbf z_\phi=(z_{0*},z_{s*},z_{q*},D_*e^{i\phi})^{\mathsf T}$,
where $D_*=|z_{d*}|$. For the present calculation,
\begin{equation}
	\mathcal F_{\mathrm{scan}}(\phi)
	=\mathcal F_{\mathrm{var}}(\mathbf z_\phi)
	=\mathcal F_{\mathrm{var,mix}}+A_\phi\cos^2\phi,
	\qquad A_\phi>0,
	\label{eq:fixed_amplitude_scan}
\end{equation}
where $A_\phi=\mathcal F_{\mathrm{scan}}(0)
-\mathcal F_{\mathrm{scan}}(\pi/2)$.
The minima at $\phi=\pm\pi/2$ correspond to the time-reversed
$s+d_{3z^2-r^2}\pm id_{x^2-y^2}$ states. The marked extrema in
Fig.~\ref{fig:feynman_GL} refer to this fixed-amplitude phase scan;
the amplitudes are not reoptimized at each phase.

In Fig.~\ref{fig:feynman_GL}, the blue solid curve shows the variational free energy at fixed mixed-state equilibrium amplitudes. The orange dashed and green dash-dotted lines represent the pure-$A_1$ and chiral pure-$E$ reference energies, respectively. The curve has degenerate minima at $\phi=\pm\pi/2$, corresponding to the time-reversed $s+d_q\pm id$ states. These minima lie below both pure-state energies, indicating an energetic preference for the mixed state over these competing states.

\section{effective odd-parity superconductivity model}

The normal state Hamiltonian with parallel spin-momentum coupling is
\begin{equation}
	H_0=\sum_\textbf{k}(c^\dagger_{\textbf{k},\uparrow},c^\dagger_{\textbf{k},\downarrow})\begin{pmatrix}\xi_\textbf{k}+\alpha_z(\textbf{k}) & \alpha_x(\textbf{k})-i\alpha_y(\textbf{k}) \\ \alpha_x(\textbf{k})+i\alpha_y(\textbf{k}) & \xi_\textbf{k}-\alpha_z(\textbf{k}) \end{pmatrix}\begin{pmatrix}c_{\textbf{k},\uparrow}\\ c_{\textbf{k},\downarrow}\end{pmatrix},
\end{equation}
where $\alpha_i(\textbf{k})=\alpha\sin k_i$ for $i=x, y, z$.
Note that $\alpha_z(\textbf{k})$ is different from the out-of-plane Zeeman field, which is used in the basis transformation mapping the spin-orbit coupling and $s+d+id$-wave pairing to an effective odd-parity pairing~\cite{kheirkhah2020first}. Here, for $k'=-k$, $\alpha_z(-\textbf{k})=-\alpha_z(-\textbf{k})$.

The even-parity superconductivity(SC) is described by
\begin{equation}
	H_\text{SC}=\sum_\textbf{k}(\Delta_s(\textbf{k})+i\Delta_d(\textbf{k}))c^\dagger_{\textbf{k},\uparrow}c^\dagger_{-\textbf{k},\downarrow}+h.c.,	
\end{equation}
where $\Delta_s(\textbf{k})=\Delta_0+\Delta_s\eta_s(\textbf{k})+\Delta_q\eta_q(\textbf{k})$ and $\Delta_d(\textbf{k})=\Delta_d\eta_d(\textbf{k})$. Here, $\Delta_{q}(\mathbf k)$ and $\Delta_{d}(\mathbf k)$
are real, even-momentum components of the $d+id$ pairing.

The SC part can be rewritten on the Nambu basis as
\begin{equation}
	H_\text{SC}=\sum_\textbf{k}(c^\dagger_{\textbf{k},\uparrow},c^\dagger_{\textbf{k},\downarrow})\begin{pmatrix}0 & -(\Delta_s(\textbf{k})+i\Delta_d(\textbf{k})) \\ (\Delta_s(\textbf{k})+i\Delta_d(\textbf{k})) & 0 \end{pmatrix}\begin{pmatrix}c^\dagger_{-\textbf{k},\uparrow}\\ c^\dagger_{-\textbf{k},\downarrow}\end{pmatrix}+h.c..	
\end{equation}

Following Ref.~\cite{alicea2010majorana,kheirkhah2020first}, we can do the transformation,

\begin{equation}
	(c^\dagger_{\textbf{k},\uparrow},c^\dagger_{\textbf{k},\downarrow})=\begin{pmatrix}
		\cos\frac{\theta_\textbf{k}}{2} & e^{-i\phi_\textbf{k}}\sin\frac{\theta_\textbf{k}}{2}\\
		e^{-i\phi_\textbf{k}}\sin\frac{\theta_\textbf{k}}{2} &-\cos\frac{\theta_\textbf{k}}{2} 
	\end{pmatrix}\begin{pmatrix}c^\dagger_{\textbf{k},+}\\ c^\dagger_{\textbf{k},-}\end{pmatrix}+h.c.,	
\end{equation}
where $\theta_\textbf{k}$ satisfies $\cos\theta_\textbf{k}=\alpha_z(\textbf{k})/\alpha(\textbf{k})$ with $\alpha(\textbf{k})=\sqrt{\alpha_x(\textbf{k})^2+\alpha_y(\textbf{k})^2+\alpha_z(\textbf{k})^2}$ and $\phi_\textbf{k}$ satisfies $e^{i\phi_\textbf{k}}=(\alpha_x(\textbf{k})+i\alpha_y(\textbf{k}))/\sqrt{\alpha_x(\textbf{k})^2+\alpha_y(\textbf{k})^2}$. This is different from
$\alpha(\textbf{k})=\sqrt{\alpha_x(\textbf{k})^2+\alpha_y(\textbf{k})^2}$ for Rashba spin-orbit coupling~\cite{kheirkhah2020first}.

The normal part can be written in a diagonal form
\begin{equation}
	H_0=\sum_\mathbf{k}(\varepsilon_\mathbf{k}+\alpha(\textbf{k}))c^\dagger_{\textbf{k},+}c_{\textbf{k},+}+(\varepsilon_\mathbf{k}-\alpha(\textbf{k}))c^\dagger_{\textbf{k},-}c_{\textbf{k},-}.
\end{equation}

Fortunately, the superconducting part can also be written in a diagonal form
\begin{equation}
	\begin{aligned}
		H_\text{SC}
		&=\frac{1}{2}\sum_{\mathbf k}\left(
		\frac{
			(\Delta_s(\mathbf k)+\Delta_q(\mathbf k)+i\Delta_d(\mathbf k))
			(\alpha_x(\mathbf k)-i\alpha_y(\mathbf k))
		}{
			\sqrt{\alpha_x^2(\mathbf k)+\alpha_y^2(\mathbf k)}
		}
		c^\dagger_{\mathbf k,+}c^\dagger_{-\mathbf k,+}
		+h.c.\right)\\
		&+\frac{1}{2}\sum_{\mathbf k}\left(
		\frac{
			(\Delta_s(\mathbf k)+\Delta_q(\mathbf k)+i\Delta_d(\mathbf k))
			(\alpha_x(\mathbf k)+i\alpha_y(\mathbf k))
		}{
			\sqrt{\alpha_x^2(\mathbf k)+\alpha_y^2(\mathbf k)}
		}
		c^\dagger_{\mathbf k,-}c^\dagger_{-\mathbf k,-}
		+h.c.\right).
	\end{aligned}
\end{equation}

Thus the Hamiltonian can be decoupled into two parts,
$H=H_++ H_-$, 
with
\begin{align}
	H_+=\sum_\textbf{k}\varepsilon_{\bf{k},+}c^\dagger_{\bf{k},+}c_{\bf{k},+}+\frac{1}{2}\sum_\textbf{k}\left(\frac{(\Delta_s(\textbf{k})+\Delta_q(\textbf{k})+i\Delta_d(\textbf{k}))(\alpha_x(\textbf{k})+i\alpha_y(\textbf{k}))}{\sqrt{\alpha^2_x(\textbf{k})+\alpha^2_y(\textbf{k})}}c^\dag_{\textbf{k},+}c^\dag_{-\textbf{k},+}+h.c.\right),\\
	H_-=\sum_\textbf{k}\varepsilon_{\bf{k},-}c^\dagger_{\bf{k},-}c_{\bf{k},-}+\frac{1}{2}\sum_\textbf{k}\left(\frac{(\Delta_s(\textbf{k})+\Delta_q(\textbf{k})+i\Delta_d(\textbf{k}))(\alpha_x(\textbf{k})+i\alpha_y(\textbf{k}))}{\sqrt{\alpha^2_x+\alpha^2_y}}c^\dag_{\textbf{k},-}c^\dag_{-\textbf{k},-}+h.c.\right).
\end{align}
where $\varepsilon_{\mathbf k,\lambda}
=\varepsilon_{\mathbf k}+\lambda\alpha(\mathbf k)$.
\end{document}